\documentclass[superscriptaddress,aps,prd,showpacs,preprint]{revtex4-1}

\usepackage{eurosym}
\usepackage{graphicx}
\usepackage{amsmath}
\usepackage{mathrsfs}
\usepackage{bm}
\usepackage{color}
\usepackage{amsmath}
\usepackage{amssymb}

\usepackage{multirow}
\usepackage{array}
\usepackage[caption=false]{subfig}

\def\be{\begin{equation}}
\def\ee{\end{equation}}
\def\bea{\begin{eqnarray}}
\def\eea{\end{eqnarray}}

\allowdisplaybreaks

\begin{document}

\title{Quasi-normal Modes of near-extremal black holes in Generalized spherically symmetric spacetime and Strong Cosmic Censorship Conjecture }
\author{Piyabut Burikham}
\email{piyabut@gmail.com}
\affiliation{High Energy Physics Theory Group, Department of Physics,
	Faculty of Science, Chulalongkorn University, Phyathai Rd., Bangkok 10330, Thailand}
\author{Supakchai Ponglertsakul}
\email{Supakchai.p@gmail.com}
\affiliation{High Energy Physics Theory Group, Department of Physics,
	Faculty of Science, Chulalongkorn University, Phyathai Rd., Bangkok 10330, Thailand}
\affiliation{Division of Physics and Semiconductor Science, Dongguk University, Seoul 04620, Republic of Korea }	
\author{Taum Wuthicharn}
\email{Taum.W5@gmail.com}
\affiliation{High Energy Physics Theory Group, Department of Physics,
	Faculty of Science, Chulalongkorn University, Phyathai Rd., Bangkok 10330, Thailand}

\date{\today }

\begin{abstract}

	A number of near-extremal conditions are utilized to simplify the equation of motion of the neutral scalar perturbations in generalized spherically symmetric black hole background into a differential equation with the P\"{o}schl-Teller potential. An analytic formula for quasinormal frequencies is obtained. The analytic formula is then used to investigate Strong Cosmic Censorship conjectures~(SCC) of the generalized black hole spacetime for the smooth initial data. The Christodoulou version of the SCC is found to be violated for certain regions of the black hole parameter space including the black holes in General Relativity while the $C^{1}$ version of the SCC is always valid.

	\vspace{5mm}
	
	
\end{abstract}

\maketitle

\section{Introduction}\label{intro}

A number of gravitational wave events has been detected by the Laser Interferometer Gravitational-Wave Observatory. The discovery of these gravitational waves confirms that the gravitational disturbances propagate through space at the speed of light \cite{Cornish:2017jml,Monitor:2017mdv}. These waves carry encoded information in their oscillatory modes unique to the sources. Since the source loses energy in the form of gravitational waves, it is described by the so-called quasinormal modes~(QNMs) where an associated frequency consists of real and imaginary part. The real part refers to frequency of oscillation while imaginary part associates with decaying or growing~(e.g. when there is superradiance) characteristic time of the wave  \cite{Kokkotas:1999bd}.  It is possible to detect the quasinormal modes in various events such as falling of massive stars into a supermassive black hole~(BH) and a merging of black holes or any highly compact objects to form a new black hole \cite{Berti:2009kk}. For merging binaries, the quasinormal modes of the merged object are found in the ring down and contain information about the mass, charge, spin, and other possible ``hairs'' of the final black hole.  It is even possible to distinguish a black hole from a wormhole and other exotics from the emitted QNMs~\cite{Kim:2008zzj, Konoplya:2016hmd, Cardoso:2016rao}.

The study of QNMs of black hole has a long story. A vast number of black holes in asymptotically flat, de-Sitter (dS) and Anti de-Sitter (AdS) spacetimes are studied in the context of QNMs (see \cite{Kokkotas:1999bd, Berti:2009kk, Konoplya:2011qq} for a well-written review on this subject). Among those black hole backgrounds, extremal black holes are interesting by their own nature. An extremal black hole possesses zero surface gravity. Thus black hole's evaporation ceases when it reaches its extremal limit. An extremal Reissner-Nordstr\"om black hole is shown to have vanishing entropy despite having non-zero horizon area \cite{Hawking:1994ii}. The stabilities of near-extremal black holes are also investigated by many authors. It is shown that extremal BTZ, charged and rotating black hole are all stable against small perturbations \cite{Crisostomo:2004hj, Richartz:2015saa}. QNMs of near-extremal BH in Weyl gravity is investigated in Ref.~\cite{Momennia:2019cfd}. Analytical study of QNMs of extremal Schwarzschild dS black hole is performed by using P\"oschl-Teller technique in Ref.~\cite{Cardoso:2003sw, MaassenvandenBrink:2003yq}. QNMs of near-extremal Schwarzschild with positive cosmological constant is numerically calculated by Yoshida and Futamase in Ref.~\cite{Yoshida:2003zz}. 

Near-extremal black hole has recently received many attentions, especially in relations to violation of the Cosmic Censorship Conjecture~(CC). The Cosmic Censorship Conjecture was proposed by Penrose~\cite{Penrose:1969pc} to ensure the independence of physics outside the horizon and the unknown singularity-regularization physics inside. After some development, the conjecture evolved into two main hypotheses namely, the Strong Cosmic Censorship~(SCC) and the Weak Cosmic Censorship~(WCC). For the Strong Cosmic Censorship, the maximal Cauchy development of initial data is inextendible and classical determinism of General Relativity is preserved. The Weak Cosmic Censorship states that if there exist singularities, they must be hidden behind horizons and cannot be extended to the future null infinity. The violation of strong cosmic censorship is closely related to the stability of the Cauchy horizon. In particular, it is shown that SCC is violated in the near-extremal regime of Reissner-Nordstr\"om~(RN) dS black hole \cite{Cardoso:2017soq, Mo:2018nnu, Dias:2018ufh}. In addition, the fate of SCC are studied in many contexts e.g. Kerr-Newman dS black hole \cite{Hod:2018lmi}, fermionic field \cite{Ge:2018vjq, Destounis:2018qnb}, non-minimally coupled scalar field \cite{Gwak:2018rba,Gwak:2019ttv} and recently in Born-Infeld dS black hole \cite{Gan:2019jac}. 

The attempt to explain an accelerated expansion of the Universe indicates that general relativity (GR) might not be the final theory of gravity. A number of alternative models of gravity have been proposed to resolve this mysterious expansion. Among many models, de Rham-Gabadadze-Tolley (dRGT) massive gravity theory \cite{deRham:2010ik,deRham:2010kj} offers a new way to describe gravity. The dRGT massive gravity is a ghost-freed, non-linear generalisation of the linear Fierz-Pauli massive gravity \cite{Fierz:1939ix} with the propagating massive spin-2 degrees of freedom. A massive graviton in dRGT theory naturally generates an effective cosmological constant \cite{Gumrukcuoglu:2011ew,Gumrukcuoglu:2011zh}, however its cosmological solutions are unstable \cite{DeFelice:2012mx}. On the other hand, various black objects have been discovered in the dRGT massive gravity theory \cite{Ghosh:2015cva,Tannukij:2017jtn,Ghosh:2019eoo,Nieuwenhuizen:2011sq}. With the presence of graviton mass term, the dRGT black holes deviate from the traditional black holes in GR. Many studies have been performed to investigate the deviation e.g. quasinormal modes \cite{Burikham:2017gdm}, greybody factor \cite{Boonserm:2017qcq} and gravitational lensing \cite{Panpanich:2019mll}.

In the previous work \cite{Ponglertsakul:2018smo}, we study the QNMs of scalar perturbation on neutral dRGT black string in the near-extremal limit with positive cosmological constant. The scalar wave equation reduces to a well-known P\"{o}schl-Teller \cite{Agboola:2008axa} form in the small universe limit~(where the event horizon approaches the cosmic horizon). An analytic formula for the quasinormal frequencies can then be obtained. In this paper, we extend our previous results by considering the near-extremal charged dRGT black holes and black strings with positive and negative cosmological constant. Not only in the near-extremal case where event horizon approaching the cosmic horizon, i.e., small universe scenario, but we also found that the approximation can be extended to the more generic cases of near-extremal black hole spacetime where the Cauchy approaching the event horizon but remotely separated from the cosmic horizon, i.e., large universe scenario.

With the information on the QNMs of generalized spacetime black hole, the SCC is investigated using criteria proposed in Ref.~\cite{Hintz:2015jkj}.  The motivation of the SCC study in generalized spacetime is to see the effect of massive gravity parameter $\gamma$~(defined below) to the structure of spacetime in relations to other physical parameters; mass, charge, and cosmological constant. Whether it would enhance or reduce the degree of the SCC violation. On theoretical basis, any classical theories of gravity should be investigated whether they would lead to violation of determinism at the classical level.  This is a very interesting topic for the study of alternative theories of gravity.  Moreover, since the generalized spacetime given by Eqn.~(\ref{gmet}) can give MOND effect to describe rotation curves of galaxies as shown in Ref.~\cite{Panpanich:2018cxo}, this promising model should also be explored whether it still lead to problem with determinism that exists in GR, and in what aspect. These are all strong motivations for the SCC study in our work.  

This paper is organized as follows. In section \ref{setup}, we explore various types of extremal black holes whose background parameters will be slightly altered to create near-extremal black hole condition. Parameter space analyses shown in Fig.~\ref{QGPSdS} and Fig.~\ref{QGPSAdS} suggest a new possibility of near-extremal black hole with small charge and large $\gamma$. It is then possible to create naked singularity out of the sub-extremal by throwing in charge particle as we demonstrate in Sect.~\ref{gedan}. The near-extremal quasinormal modes of charged black holes in dS and AdS massive gravity background are analytically solved in section \ref{Ana}. Implications for SCC are analyzed in Section \ref{SCCSec}. Analog black string is discussed in Sect.~\ref{bstring}. Section \ref{conclude} concludes our work. Appendix~\ref{ReS} lists rescaling relation of spacetime parameters relevant for exploration of parameter space in generic region. Appendix~\ref{Rproof} proves certain property of surface gravity used in the SCC analysis. Appendix~\ref{SCCP} contains proof of existence and position of gaps in the lower and upper bounds used in consideration of the SCC violation.

\section{Various Types of Near-Extremal Black Hole in Massive Gravity Background}\label{setup}

The spacetime of charged black hole in dRGT massive gravity is given by \cite{Ghosh:2015cva}
\be
ds^{2}=-f(r)dt^{2}+f^{-1}(r)dr^{2}+r^{2}d\theta^{2}+r^{2}\sin^{2}\theta~d\varphi^{2},
\ee
where
\begin{equation}\label{gmet}
		f(r)=(1+\epsilon_{0})-\frac{2M}{r}+\frac{Q^{2}}{r^{2}}-\frac{\Lambda}{3}r^{2}+\gamma r.
\end{equation}
The mass and charge of the black hole are denoted by $M$ and $Q$. The cosmological constant $\Lambda$, the parameter $\gamma$ and $\epsilon_{0}$ are naturally generated from a graviton mass in dRGT massive gravity. The $\gamma$ term is associated with universal acceleration of the test particle~(similar effect to MOND~\cite{Panpanich:2018cxo}) moving in the vicinity of the black hole. For simplicity, we also set $\epsilon_{0}=0$ in this work~\cite{Kareeso:2018xum}. We can also take this form to be the generalized spherically symmetric metric of the black hole spacetime and study it phenomenologically, see e.g. Ref.~\cite{Panpanich:2018cxo}. 

The metric function (\ref{gmet}) has four possible roots. Each real positive root corresponds to a horizon of the black hole. For an asymptotically de Sitter~(dS) spacetime, where $\Lambda>0$, the roots of $f(r)$ are defined as $r_{-}$, $r_{C}$, $r_{H}$ and $r_{\Lambda}$. The $r_{-}$ has a negative value, which is unphysical. The $r_{C}$ is the Cauchy horizon, also referred to as the inner horizon which covers the singularity. The $r_{H}$ is the event horizon and $r_{\Lambda}$ is the cosmological horizon. For an asymptotically AdS spacetime, where $\Lambda<0$, the possible roots of $f(r)$ are defined as $r_{C}$, $r_{H}$, $r_{\Lambda-}$ and $r_{\Lambda+}$. An extremal black hole is a black hole with two horizons coincide into one. The surface gravity at the extremal horizon vanishes implying zero Hawking temperature. However, extremal horizon only occurs at precise value of spacetime parameters. Slightly off-setting would result in a near-extremal black hole or a naked singularity. The main focus of this work is the near-extremal black holes with nonzero cosmological constant.

\begin{figure}[h]
		\centering
		\subfloat[]{\includegraphics[width=0.41\textwidth]{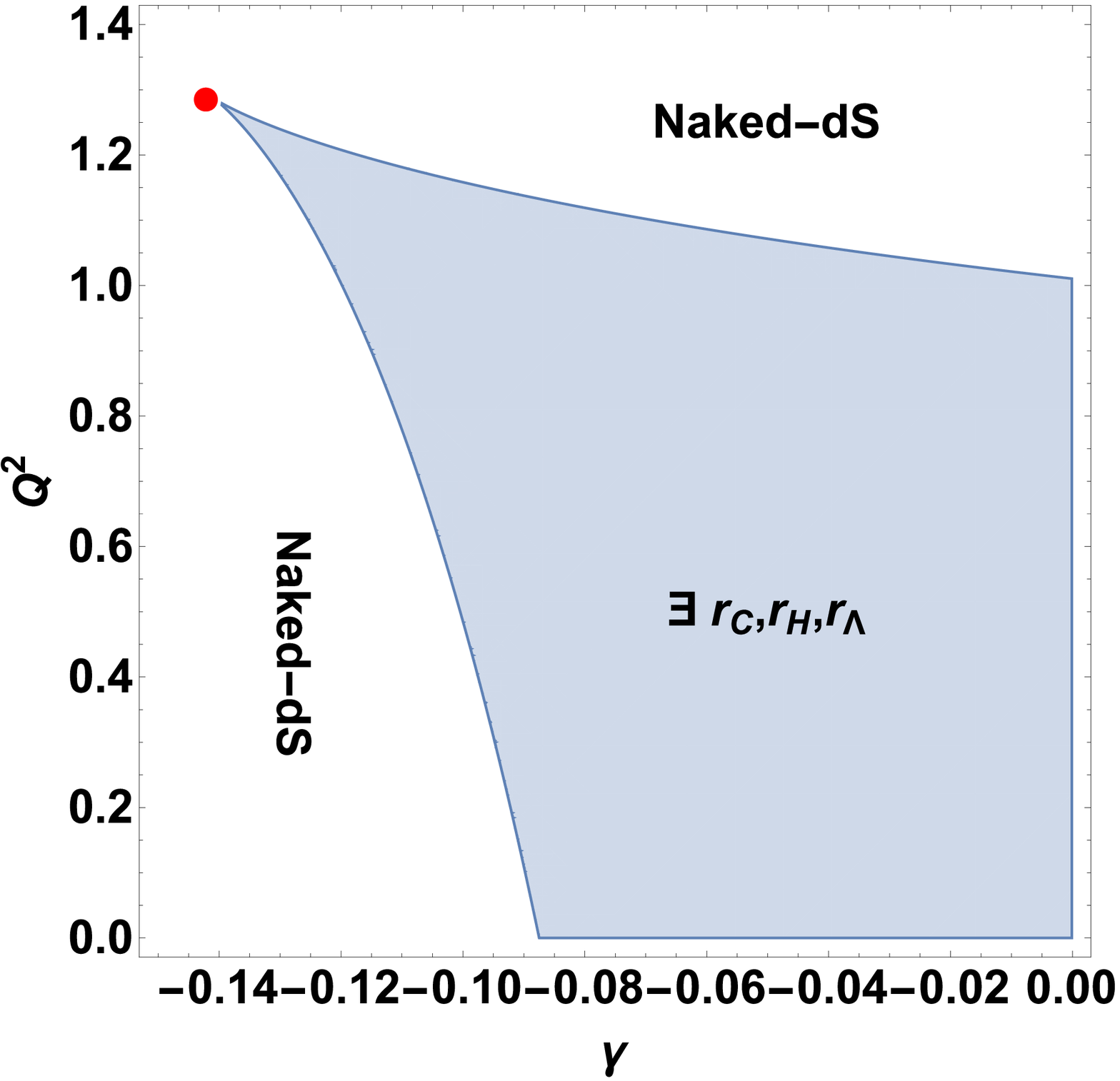}}
		\hspace{2em}
		\subfloat[]{\includegraphics[width=0.4\textwidth]{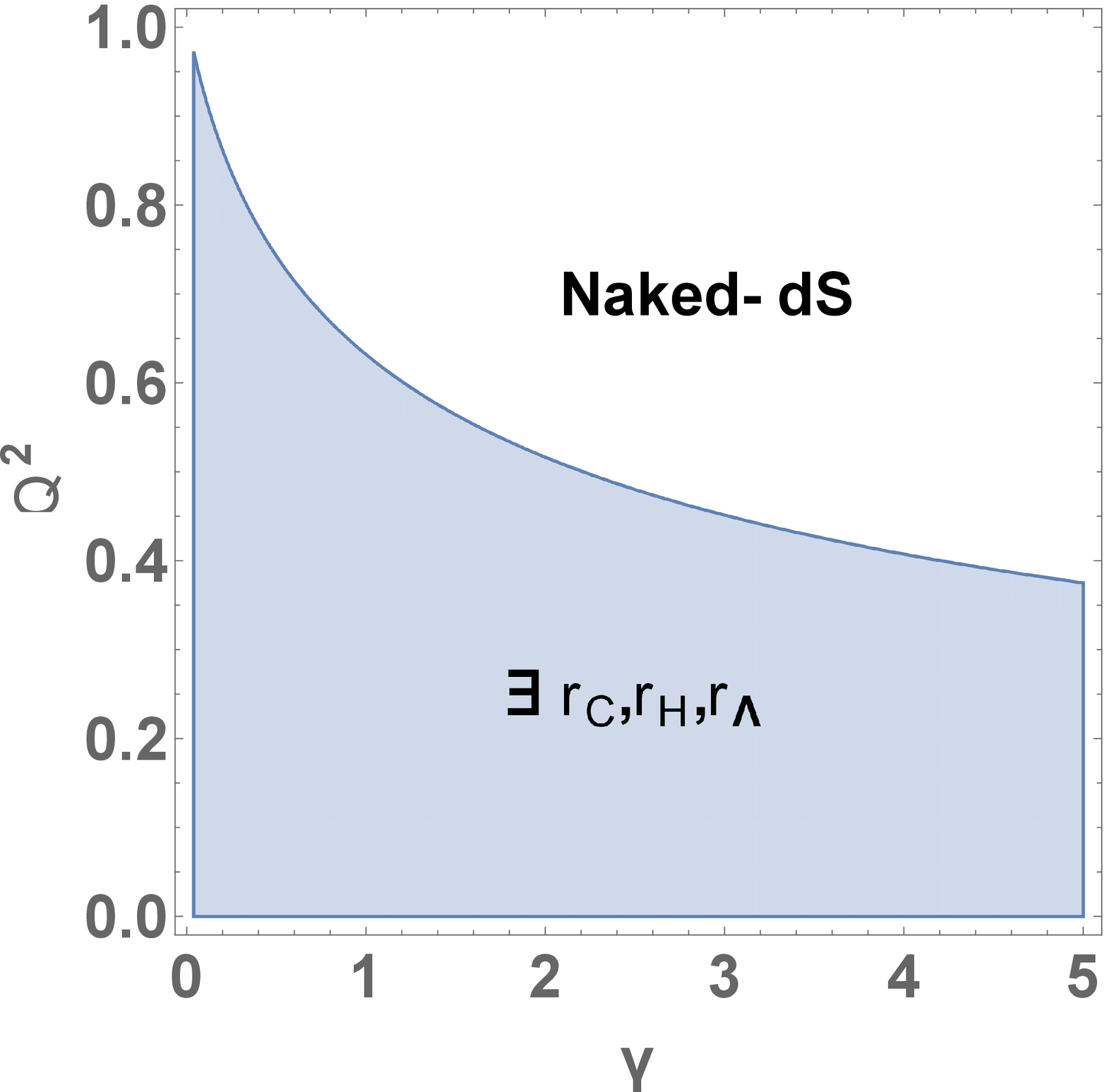}}
		\caption{A $\gamma$-$Q^{2}$ parameter space with $M=1$, $\epsilon_{0}=0$ and $\Lambda=0.03$. (Left) parameter space of black holes with $\gamma<0$ has two distinct regions. (Right) parameter space of black holes with $\gamma>0$ has two distinct regions where the naked-dS region is of the same type as the upper region of the left figure. }		\label{QGPSdS}
	\end{figure}

\begin{figure}[h]
		\centering
		\subfloat[]{\includegraphics[width=0.327\textwidth]{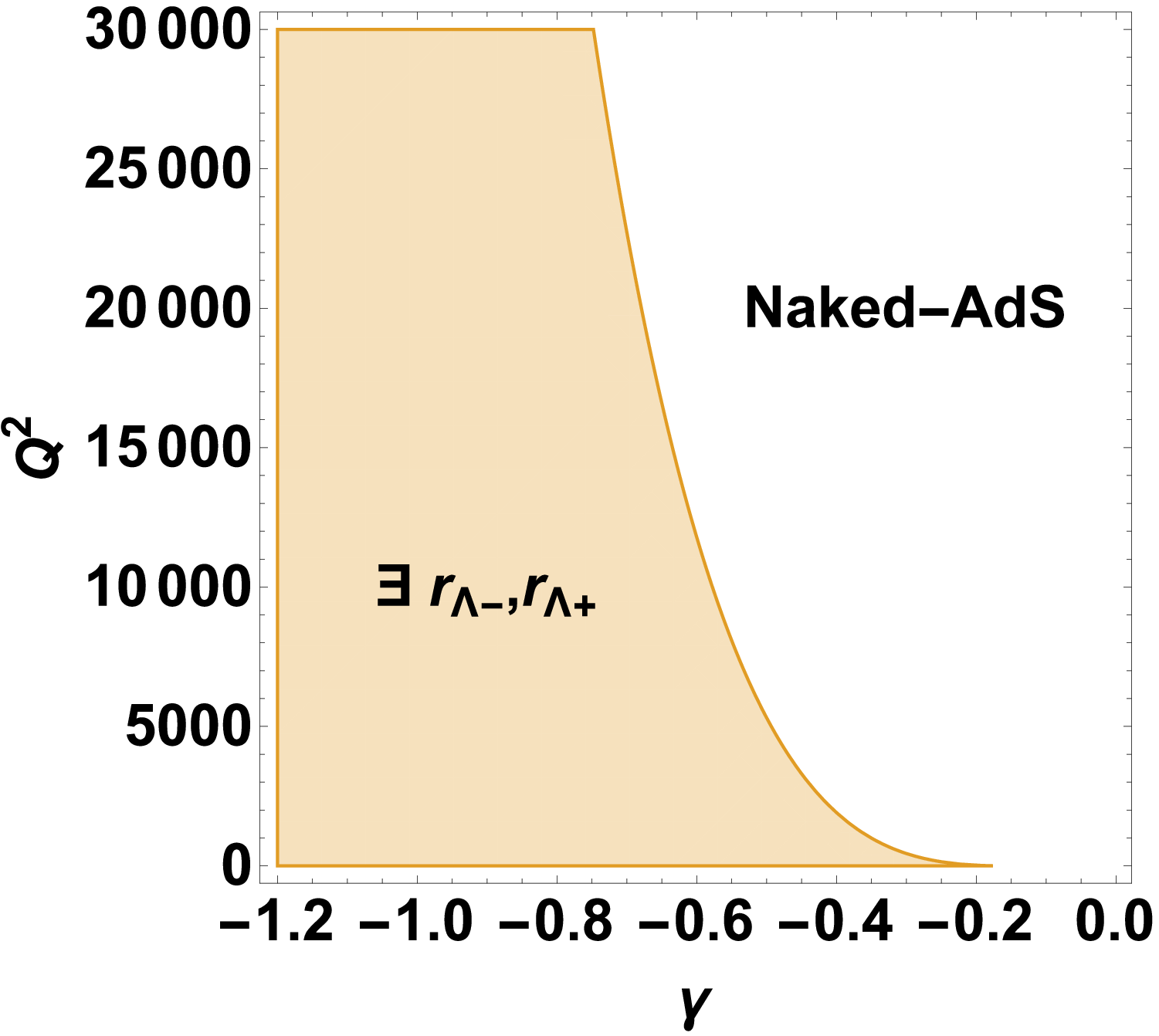}}
		\hspace{1em}
		\subfloat[]{\includegraphics[width=0.3\textwidth]{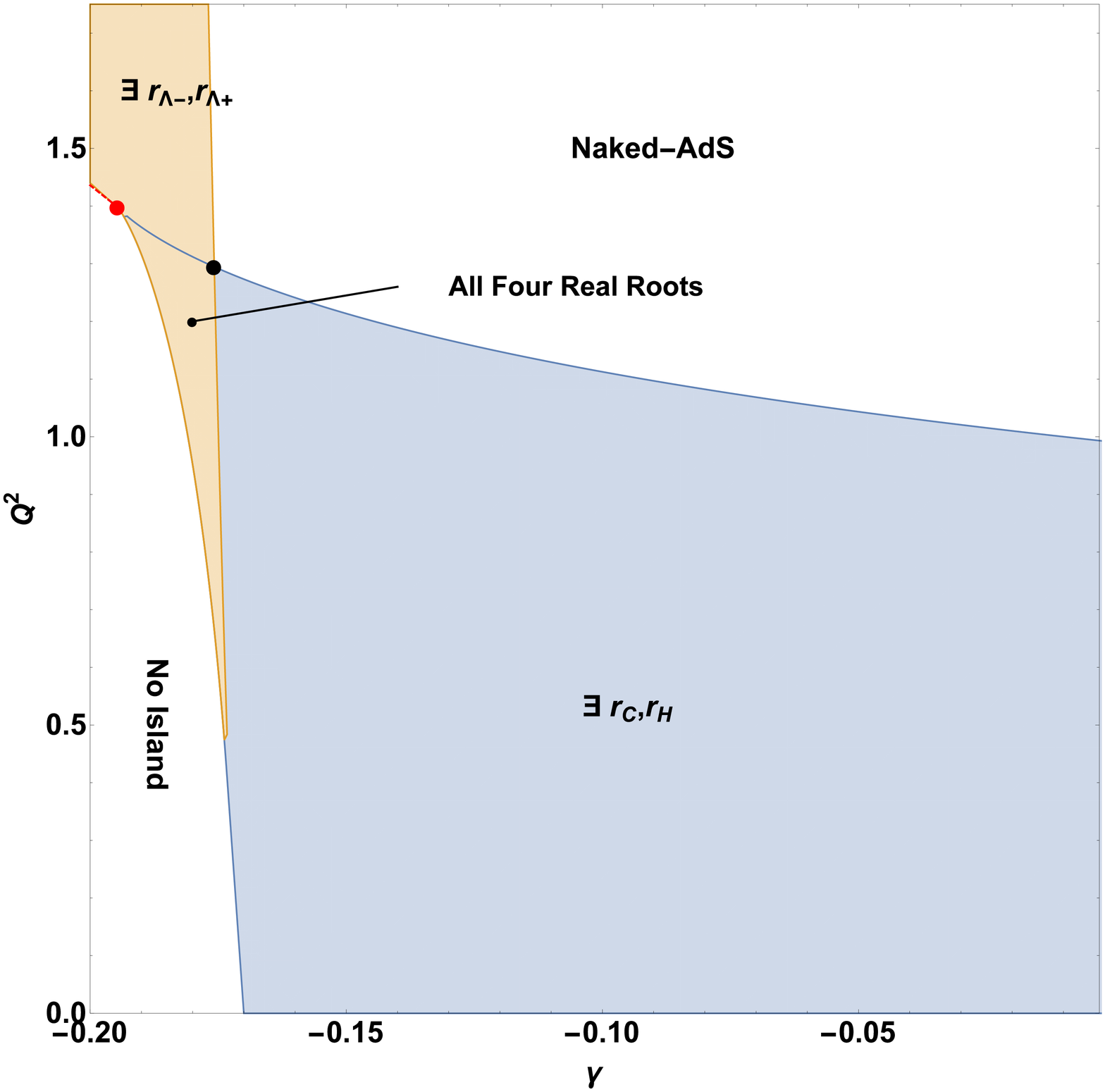}}
		\hspace{1em}
		\subfloat[]{\includegraphics[width=0.285\textwidth]{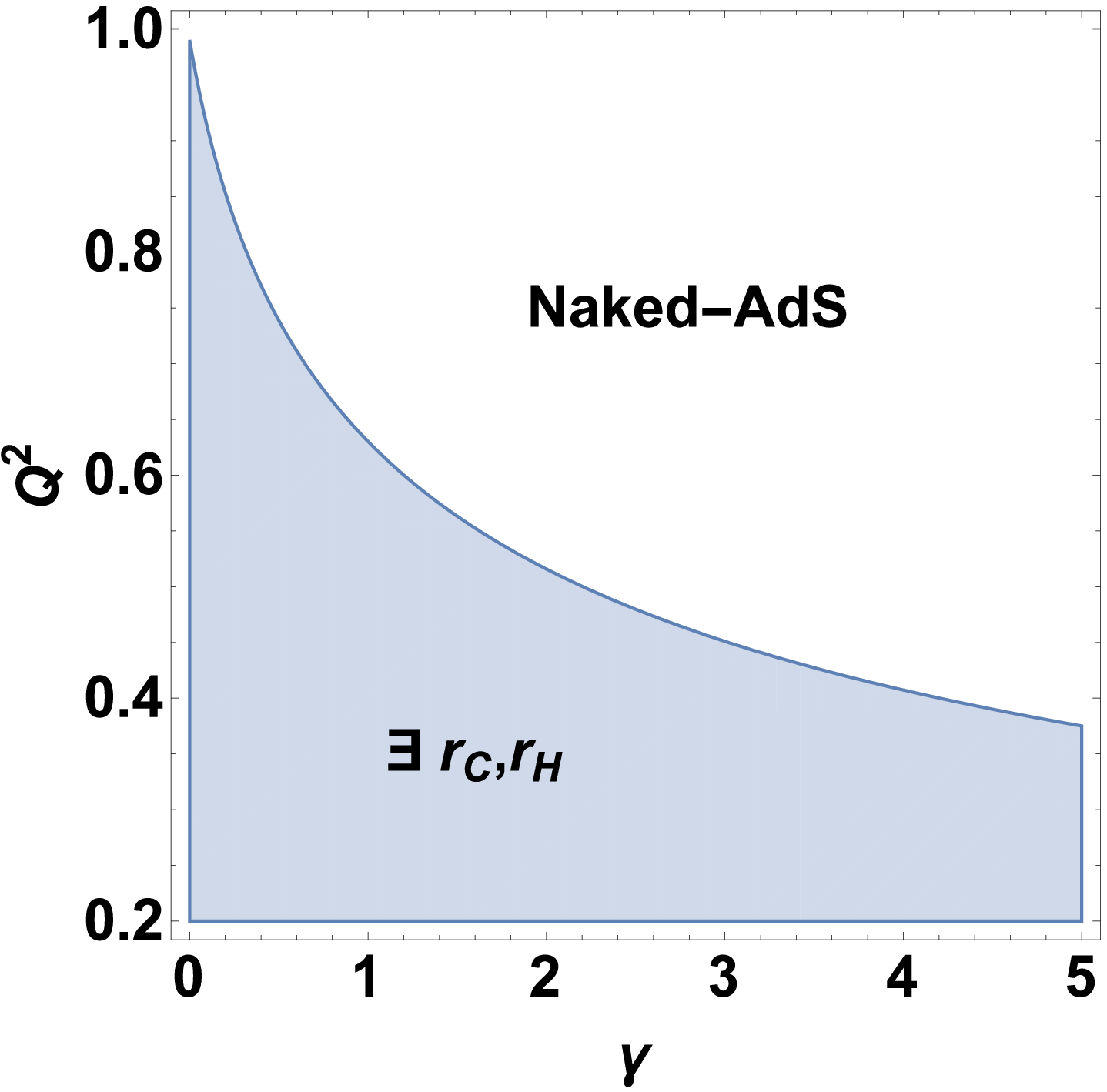}}
		\caption{Parameter spaces of black holes with $M=1,\epsilon_{0}=0$ and $\Lambda=-0.03$. (Left) parameter space of black holes with two horizons, $r_{\Lambda_-}$ and $r_{\Lambda_+}$ and ``Naked-AdS'' refers to a region where $f(r)$ has no real roots.	(Center) parameter space of the black holes with a smaller $Q^2$ where various scenarios occur, red dot is the end of two extremal lines and black dot is the double-extremal point. Dashed red line is not an extremal line. (Right) Extended parameter space with positive $\gamma$.	}
		\label{QGPSAdS}
	\end{figure}

In this work, we shall specifically focus on the near-extremal scenario with nonzero cosmological constant. In addition, we choose to focus our investigation to the region with small cosmological constant, $-0.03<\Lambda<0.03$, since this choice will cover both positive and negative values of $\Lambda$.  Additionally, spacetime with extremely small cosmological constant is strongly prefered by astrophysical observations so it deserves to be explored prior to other regions of the parameter space. We have also scanned the parameter space with larger $|\Lambda|$ and verified that the parameter space do not change significantly, each parametric region in Fig.~\ref{QGPSdS} and Fig.~\ref{QGPSAdS} simply moves slightly in the parameter space but otherwise looks similar.

\begin{figure}[h]
		\centering
		\subfloat[]{\includegraphics[width=0.4\textwidth]{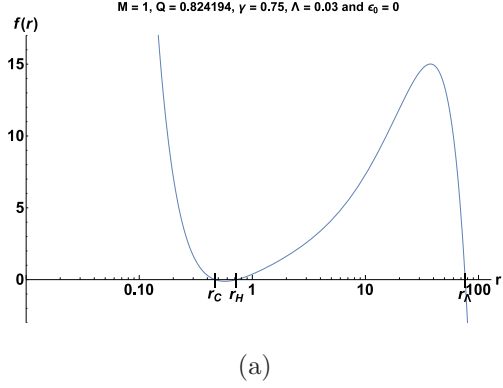}}
		\hspace{2em}
		\caption{Behaviour of $f(r)$ in near-extremal cases with positive cosmological constant for $r_{C}\sim r_{H}$. This scenario has large universe in $r_{H}<r<r_{\Lambda}$ region. }
		\label{EEdS}
	\end{figure}

We shall now analyze the roots of $f(r)$ in asymptotically dS case. The parameter space of $Q^{2}$ and $\gamma$ is shown in Fig. \ref{QGPSdS}. The $\exists~r_{C},r_{H},r_{\Lambda}$ is a region where $f(r)$ has three positive real roots, thus black holes in this region possess three distinct horizons. The region labeled as Naked-dS is a region where black hole has only one (cosmological) horizon. The boundary between $\exists~r_{C},r_{H},r_{\Lambda}$ and the upper naked-dS region in Fig. \ref{QGPSdS}a is the extremal case where $r_C=r_H$. Another extremal case with $r_H=r_{\Lambda}$ is illustrated as the boundary of the shaded area and the lower naked-dS region in the lower-left corner of Fig. \ref{QGPSdS}a. The near-extremal conditions can be obtained by considering the neighborhood area under and over the extremal lines. We observe that the case $r_H=r_{\Lambda}$, which yields a very small physical spacetime region, vanishes for $\gamma>0$ as shown in Fig.~\ref{QGPSdS}b. The red dot at $(\gamma,Q^{2})=(-0.140927,1.28693316)$ is the end point of two extremal lines. The notable case of near-extremal cases are shown in Fig. \ref{EEdS} where the physical universe is vast.


For asymptotically AdS case with $\Lambda<0$, $f(r)$ has four positive real roots, $r_{C}$, $r_{H}$, $r_{\Lambda-}$ and $r_{\Lambda+}$. The $r_{C}$ and $r_{H}$ are Cauchy horizon and event horizon respectively which are similar to their counterparts in asymptotically de-Sitter space. The $r_{\Lambda-}$ acts as if it is $r_{\Lambda}$ in asymptotically de-Sitter space. The physical Universe bounded between $r_{H}$ and $r_{\Lambda-}$ behaves locally similar to the physical Universe from asymptotically de-Sitter space. The $r_{\Lambda+}$ plays a similar role to the event horizon for another physical universe. This universe is asymptotically AdS and possesses only one horizon.

\begin{figure}[h]
		\centering  
		\subfloat[]{\includegraphics[width=0.4\textwidth]{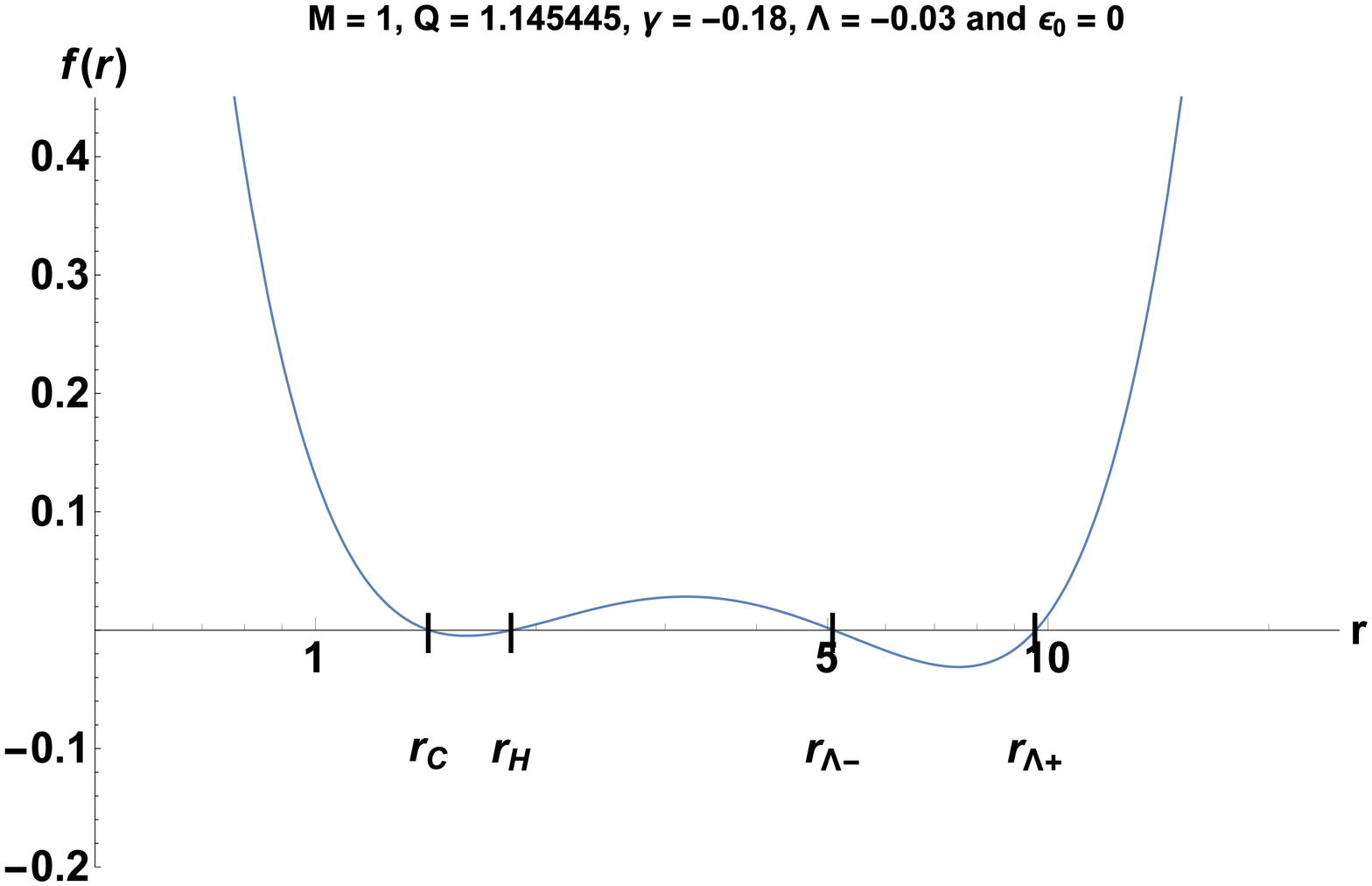}}
		\hspace{2em}
		\subfloat[]{\includegraphics[width=0.4\textwidth]{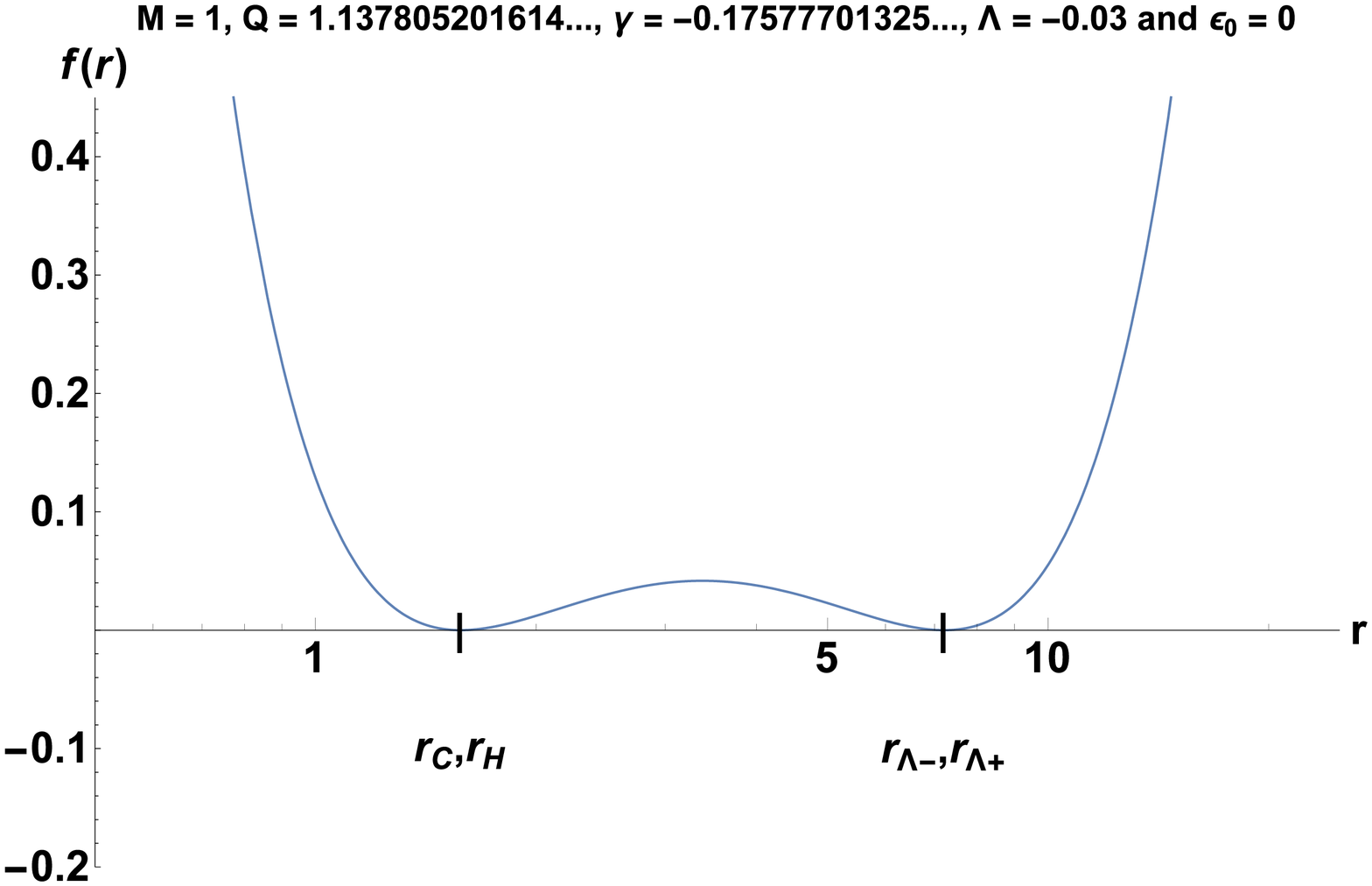}}
		\caption{Examples of $f(r)$ from near-extremal black hole in asymptotically Anti de-Sitter space. (Left) The near-extremal black hole of the type $r_{C}\sim r_{H}$. (Right) The double (near-)extremal black hole of the type $r_{C}=r_{H}$ and $r_{\Lambda-}=r_{\Lambda+}$.}
		\label{EEAdS3}
	\end{figure}

The numerical roots of $f(r)$ with negative cosmological constant are investigated in $Q^{2}$-$\gamma$ parameter space as shown in Fig. \ref{QGPSAdS}. In Fig. \ref{QGPSAdS}a, the shaded area denotes the black holes with two positive real roots, $r_{\Lambda-}$ and $r_{\Lambda+}$. The naked-AdS area refers to the case where $f(r)$ has no real root. The boundary between these areas is the extremal case for which $r_{\Lambda-}=r_{\Lambda+}$. Below this extremal line is the near-extremal limit for black holes with $r_{\Lambda-}\sim r_{\Lambda+}$. For a smaller region of $Q^2$ and negative $\gamma$, various scenarios exist in this region as shown in Fig. \ref{QGPSAdS}b. The shaded area where $\exists~r_{\Lambda_-},r_{\Lambda_+}$ and $\exists~r_{c},r_{H}$ overlap, black holes with all four positive roots exist. At the lower left corner of the plot, there exists an area with only $r_C$ and $r_{\Lambda_+}$ exist. This is denoted by ``No Island''. The red dot at $(\gamma,Q^{2})=(-0.1946299549265,1.3981644129925)$ is the end point of two extremal lines, $r_{H}=r_{\Lambda-}$ and $r_{H}=r_{C}$. The $\exists~r_{\Lambda_-},r_{\Lambda_+}$ region can be extended further towards increasing $Q^2$ direction and joins smoothly with the parameter space display in Fig. \ref{QGPSAdS}a. The $\exists~r_{C},r_{H}$ region represents an area with no $r_{\Lambda_-},r_{\Lambda_+}$ roots. The horizonless spacetime is labeled as ``Naked-AdS'' region. The $\exists~r_{C},r_{H}$ and Naked-AdS regions exist for non-negative $\gamma$ as shown in Fig. \ref{QGPSAdS}c. The near-extremal black hole spacetime can be found by choosing the parameters that are very close to the lines separating each region. Examples of near-extremal black holes with $r_{C}\sim r_{H}$ and $r_{\Lambda_-}\sim r_{\Lambda_+}$ are displayed in Fig. \ref{EEAdS3}.
	
There are two more interesting scenarios in Fig. \ref{QGPSAdS}b. The line where ``No Island'' region meets with ``All Four Real Roots'' region represents an extremal case where $r_{H}=r_{\Lambda-}$. The small region on the right hand side of this line is a near-extremal condition of the $r_{H}\sim r_{\Lambda-}$ type. By manipulating parameters in the theory, it is possible to find a black hole with double extremal horizons i.e., $r_{C}=r_{H}$ and $r_{\Lambda-}=r_{\Lambda+}$. This can be found at the intersection between three regions i.e.,  All Four Real Roots, $\exists~r_{C},r_{H}$ and $\exists~r_{\Lambda_-},r_{\Lambda_+}$. The $r_{C}\simeq r_{H}$ and double extremal cases are shown in Fig. \ref{EEAdS3}, they are relevant in our SCC consideration below.

\subsection{An exotic near-extremal black hole and Creation of over-extremal spacetime, a naked singularity}\label{gedan}

From the analysis of $\gamma$-$Q^{2}$ parameter space in Fig.~\ref{QGPSdS}b and Fig.~\ref{QGPSAdS}c, it is found that the extremal of the type $r_{C}=r_{H}$ requires a very small amount of $Q^{2}$ once the value of $\gamma$ becomes rather large. This property would reduce any back reaction from the charged particle being dropped into the black hole. It follows that a near-extremal black hole would be able to absorb greater range of charged particle in comparison to near-extremal in other background. As such, it might be possible to create an over-extremal spacetime from a sub-extremal black hole in massive gravity background. This process of WCC violation has been highly discussed in massless gravity theory. 

In standard GR, it was originally argued by Wald \cite{ Wald1974} that the event horizon of extremal Kerr-Newman black hole cannot be destroyed by a test particle. However, WCC can be evaded if the black hole is initially near-extreme and the effect of the back-reaction of the particle is neglected. Starting with near-extremal cases, RN \cite{Hubeny:1998ga} and Kerr \cite{Jacobson:2009kt} black hole are found to expose sigularity by absorbing a test particle. More interestingly, Kerr-Newman black hole can be overspun in an extremal~(with negative $\Lambda$) \cite{ Zhang2013} and near-extremal \cite{ Saa:2011wq} scenarios by using an electrically charged particle with angular momentum. In the presence of $\Lambda$, extremal RN-AdS and extremal Kerr-dS/AdS black holes can be overcharged and overspun \cite{ Zhang:2013tba}. The near-extremal Kerr-Newman-AdS black hole violates WCC by absorbing test particle into its event horizon \cite{Song:2017mdx}. When quantum mechanical tunneling is taken into account, the event horizon of nearly extremal rotating black hole \cite{ Richartz:2008xm} and charged black hole \cite{ Richartz:2011vf} are destroyed by quantum absorption of a fermionic test particle into the black hole. In contrast, Sorce and Wald \cite{ Sorce:2017dst} have recently shown that when the full second order correction to the black hole mass and self-force effect are considered, a Kerr-Newman black hole cannot be overcharged/overspun. This work has put all the Hubeny type of (classical) gadanken experiments~\cite{Hubeny:1998ga} into question. And the violation of WCC remains as an unsettled problem.

In conventional GR at zero spin, a test mass needs to carry charge that is over extremal, $q^{2}>m$ by itself in order to make the total charge of the RN black hole to exceed the extremal limit $Q^{2}>M$. In our case the difference is the additional $\gamma$ parameter of the black hole background that allows a test mass with sub-extremal charge to be absorbed and results in the naked singularity spacetime. Remarkably, additional gravitational pull from the $\gamma$ term provides exotic possibility of near-extremal black hole where charge is far away from extremality $Q^{2}\ll M$, the new kind of near-extremal black hole which does not exist in GR. This kind of near-extremal black hole can thus absorb test particle with sub-extremal charge and still results in the naked singularity.

	\begin{figure}[h]
		\centering
		\subfloat[]{\includegraphics[width=0.4\textwidth]{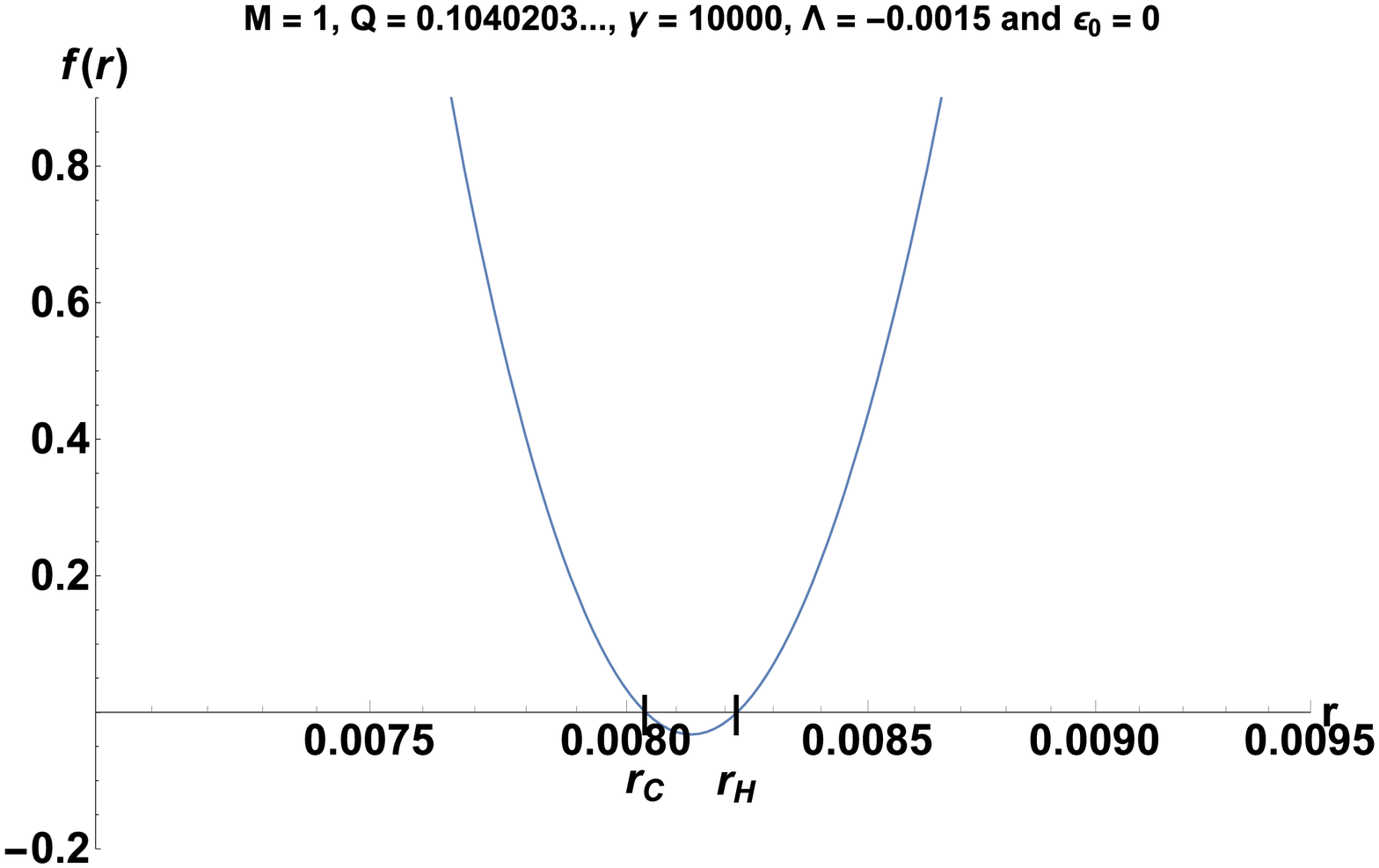}}
		\hspace{2em}
		\subfloat[]{\includegraphics[width=0.4\textwidth]{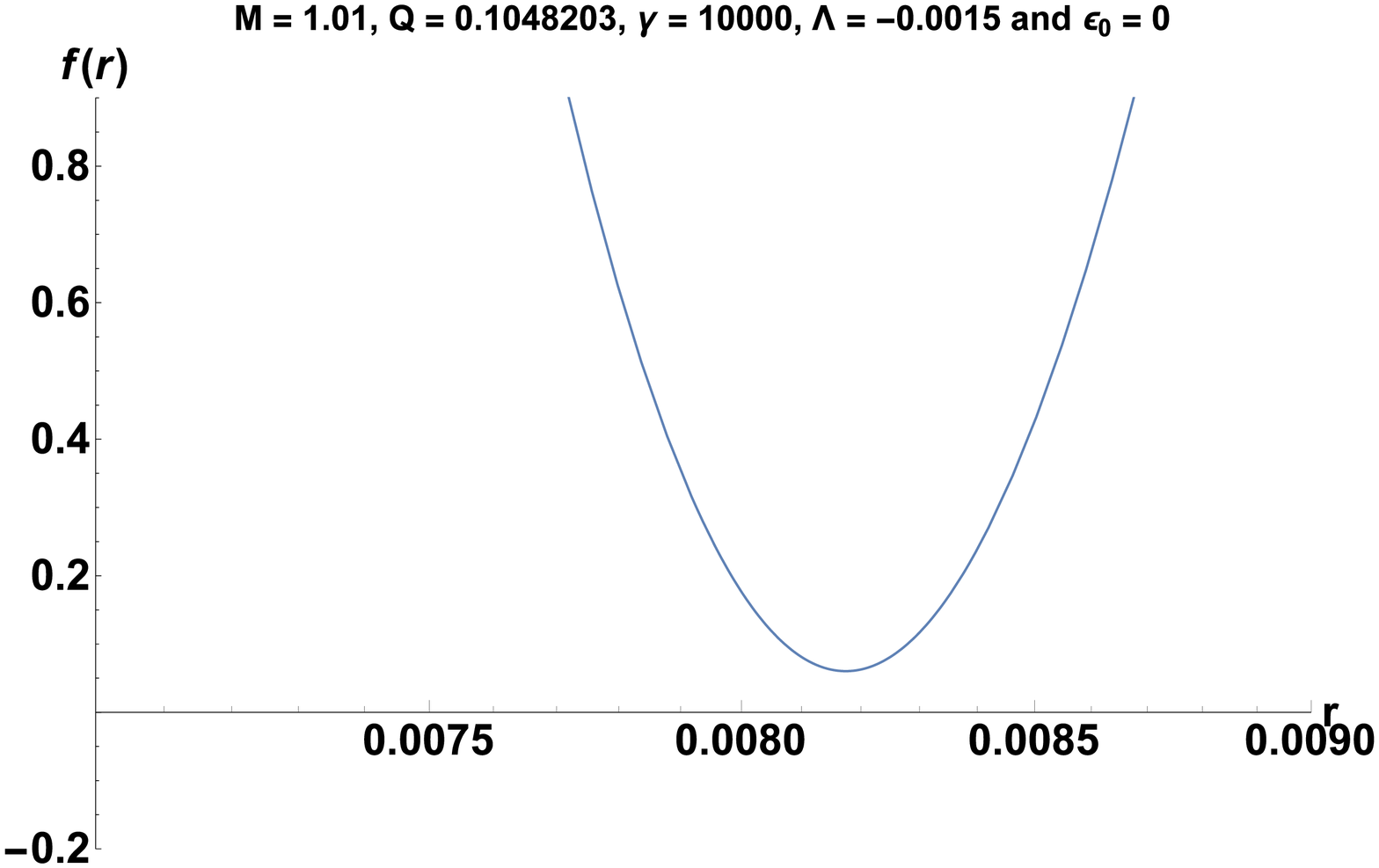}}
		\caption{An example of an over-extremal black hole created from a sub-extremal black hole in massive gravity background. The high-value $\gamma$ allows the possibility of the sub-extremal charged particle being absorbed by the sub-extremal black hole. (Left) The sub-extremal black hole before absorbing the charged particle. (Right) The over-extremal black hole after absorbing the charged particle.}
		\label{StoO}
	\end{figure}
One example of such process is shown in Fig. \ref{StoO}. The black hole is set with the following parameter, $M=1$ and $Q=0.1040203$, while the spacetime parameters are $\epsilon_{0}=0, \gamma=10,000$ and $\Lambda=-0.0015$. The absorbed charged particle possesses mass $m=0.01$ and charge $q=0.0008$ with sub-extremal charge/mass ratio. Although both charges of the black hole and the test particle are quite small, the size of event horizon is also tiny. The test particle with sub-extremal $q/m < 1$ only needs to be injected with sufficient kinetic energy to overcome the Coulomb repulsion at the near-extremal horizon. Once the test particle passed the horizon, the spacetime should become over-extremal and the singularity at $r=0$ could be exposed, at least momentarily if not permanently. The question becomes whether the spacetime fluctuations around the near-extremal horizon(s)~(Cauchy and event horizons very close together) due to the absorption of test particle would result in a stable over-extremal spacetime with naked singularity or end up as a sub-extremal black hole spacetime after emission of energy through the QNM oscillations. We leave this interesting problem for future investigation. 

\section{Analytic Solution of Quasinormal Modes for Various Near-Extremal Charged Black Holes in Massive Gravity Background}\label{Ana}

First we write down the wave equation of neutral scalar field in the generalized spherically symmetric background of the near-extremal charged black holes, $\Box \Phi(x) =0$. The scalar field $\Phi(x)$ can be expressed as $\Phi(\vec{r},t) = \frac{\phi(r)}{r}Y_{l}^{m}(\hat{r}) e^{-i\omega t}$ with $Y_{l}^{m}$ the spherical harmonics. The radial wave equation of $\phi(r)$ then takes the form
	\begin{equation}
		\frac{d^{2}\phi}{dr^{2}_{*}}+\left[\omega^{2}-f(r)\left(m^{2}_{s}+\frac{l(l+1)}{r^{2}}+\frac{f'(r)}{r}\right)\right]\phi=0, \label{KGeq}
	\end{equation}
where the tortoise coordinate $r_{*}$ is defined by $\displaystyle{\frac{dr_{*}}{dr}=\frac{1}{f}}$. Mass of the scalar field is denoted by $m_s$. $l, m$ are the azimuthal angular quantum number and magnetic quantum number respectively. With the form of $f(r)$ given by (\ref{gmet}), the metric function has four roots regardless of whether they are real or complex numbers. We define these four roots as $A$, $B$, $C$ and $D$ which can be regarded as horizon if its value is a real positive number. For example, if we choose parameter in the ``All Four Real Roots'' region in Fig. \ref{QGPSAdS}, then we can associate $(A,B,C,D)$ with $(r_C,r_H,r_{\Lambda_-},r_{\Lambda_+})$. The metric can then be rewritten as
	\begin{equation}\label{gmet1}
		f(r)=-\frac{\Lambda}{3r^{2}}(r-A)(r-B)(r-C)(r-D).
	\end{equation}
Let's suppose that $A$ is one of the black hole horizon which corresponding surface gravity is positive i.e., $\kappa_A \equiv \frac{f'(A)}{2} > 0$. Therefore the surface gravity at the horizon $A$ is 
	\begin{eqnarray}
		\kappa_{A}&\equiv&\frac{1}{2}\left.\frac{df}{dr}\right|_{r=A}=-\frac{\Lambda/3}{2A^{2}}(A-B)(A-C)(A-D)=\frac{\Pi}{2A^{2}}(A-B),
	\end{eqnarray}
where $\Pi=\displaystyle{-\frac{\Lambda}{3}(A-C)(A-D)}$. By comparing the two expressions of $f(r)$, $(\ref{gmet})$ and $(\ref{gmet1})$ and considering the near-extremal limit $B\to A$, the following relations are obtained
\begin{align}
		C+D &\sim\frac{3\gamma}{\Lambda}-2A, 
		~CD  \sim -\frac{3}{\Lambda}\left(\frac{Q}{A}\right)^{2},\nonumber \\
		\frac{\Lambda}{3} &= \frac{(A+B)\gamma+\tau-Q^{2}/AB}{(A^{2}+B^{2}+AB)}\sim\frac{(2A\gamma+\tau)-Q^{2}/A^{2}}{3A^{2}},\nonumber 
\end{align}
where $\tau=1+\epsilon_{0}$.
Then we expand the following term around $A$,
	\begin{eqnarray}
		(r-C)(r-D)&=&(A-C)(A-D)+(2A-C-D)(r-A)+O(r-A)^{2},\nonumber \\
		&\equiv&(A-C)(A-D)\Big[1+p(r,A,C,D)\cdot(r-A)\Big],\nonumber
	\end{eqnarray}
where $p(r,A,C,D)$ is some function of $A$, $C$ and $D$. By adding the next leading order of $(r-A)$, the above expansion is valid in the range further away from the extremal horizon. This expansion is performed to prevent our solution from overlapping with the near-horizon modes discussed in Ref.~\cite{Tangphati:2018jdx}, which is another generic class of QNMs of black hole. While the QNMs we calculate in this work are the all-region modes that can propagate through the whole physical spacetime region~(i.e., WKB modes), the near-horizon modes only exist close to the corresponding horizons.  It is interesting that in the near-extremal case, the two kinds of QNMs coexist along the same imaginary axis for neutral scalar perturbations.  

Using the above expansion, the tortoise coordinate can be expressed as 
	\begin{eqnarray}
		dr_{*}&=&-\frac{3}{\Lambda}\frac{r^{2}dr}{(r-A)(r-B)(r-C)(r-D)},\nonumber \\
			&=&-\frac{3}{\Lambda}\frac{dr}{(A-C)(A-D)}\left\lbrace\frac{1}{A-B}\left(\frac{A^{2}}{r-A}-\frac{1}{1-p\cdot(A-B)}\frac{B^{2}}{r-B}\right)\right.\nonumber \\
		&&\left.+\frac{(Ap-1)^{2}}{1-p\cdot(A-B)}\frac{1}{1+p\cdot(r-A)}\right\rbrace.\nonumber
	\end{eqnarray}
With the near-extremal condition~($A\sim B$), the above equation can be approximated~(the last term in $\{...\}$ is subdominant with respect to terms of order $\mathcal{O}((A-B)^{-1})$) as,
	\begin{eqnarray}
		C_1e^{2\kappa_{A}r_{*}}&\sim&(r-A)(r-B)^{-(B/A)^{2}},\nonumber
	\end{eqnarray}
where $C_1$ is integration constants. It is therefore possible to express the radial coordinate $r$ in term of tortoise coordinate $r_\ast$
	\begin{align}
		C_1e^{2\kappa_{A}r_{*}}&\sim\frac{r-A}{r-B},&\nonumber \\
		r&\sim\frac{C_1Be^{2\kappa_{A}r_{*}}-A}{-1+C_1e^{2\kappa_{A}r_{*}}}=\frac{Be^{2\kappa_{A}r_{*}}+A}{1+e^{2\kappa_{A}r_{*}}}.\label{rtor}
	\end{align}
where $C_1$ is chosen to be $-1$ in the last step. 

Notice that the above approximation only requires that the physical Universe is not bounded by the extremal horizons on both sides. In the case where the physical Universe is bounded by the extremal horizons on both sides, however, the following relationship is obtained, $A>r>B$. Since $A\sim B$, the value of $r$ can be approximated to $A$ unless $(r-A)(r-B)$ term is concerned. The tortoise coordinate can be expressed as
	\begin{eqnarray}
		dr_{*}&\cong&\frac{A^{2}}{\Pi}\int\frac{dr}{(r-A)(r-B)},\nonumber \\
		\tilde{C}_1 e^{2\kappa_{A}r_{*}}&=&\frac{r-A}{r-B},\nonumber
	\end{eqnarray}
for constant of integration $\tilde{C}_1$. This would inevitably lead to the same result for $\tilde{C}_1= -1$. Thus Eqn.~(\ref{rtor}) is valid regardless of the type of extremality. 

From the above relations, the metric function may be expressed in the tortoise coordinate as,
	\begin{align}
		f(r_{*})&\sim \frac{(A\kappa_{A})^{2}}{(A\gamma+\tau-2Q^{2}/A^{2})\cosh^{2}(\kappa_{A}r_{*})}.
	\end{align}
Thus, the radial Klein-Gordon equation (\ref{KGeq}) becomes,
	\begin{equation}
		\frac{d^{2}\phi}{dr^{2}_{*}}+\left[\omega^{2}-\frac{V_{0}(A)}{\cosh^{2}(\kappa_{A}r_{*})}\right]\phi=0,  \label{reom}
	\end{equation}
where,
	\begin{equation}
		V_{0}(r)=\frac{\kappa^{2}(r)}{(r\gamma+\tau-2Q^{2}/r^{2})}\Big[m^{2}_{s}r^{2}+l(l+1)\Big].
	\end{equation}
The potential is the well-known P\"{o}schl-Teller potential \cite{Poschl1933}. By applying the boundary condition of quasinormal modes, the following associated quasinormal frequencies are obtained \cite{Agboola:2008axa},
	\begin{equation}\label{FM}
		\omega_{n}=\kappa_{A}\left\lbrace\sqrt{\frac{V_{0}(A)}{\kappa^{2}_{A}}-\frac{1}{4}}-\left(n+\frac{1}{2}\right)i\right\rbrace.
	\end{equation}
The solution is obtained with only requirement for $A$ and $B$ to be real number. $C$ and $D$ can be either complex or real numbers without changing the result of our calculation.

\subsection{Solutions in Asymptotically dS Background}

In the vicinity of event horizon and cosmological horizon, general solution of (\ref{KGeq}) can be expressed as
\begin{align}
\phi_{in} \sim \left\{ \begin{array}{lr}
  e^{-i\omega r_{\ast}},  \hspace{3.1cm} \mbox{ as $r\rightarrow r_H$} \\
  C'_1 e^{-i\omega r_{\ast}} + C'_2 e^{i\omega r_{\ast}}, \hspace{1.1cm} \mbox{ as $r\rightarrow r_{\Lambda}$}.
       \end{array} \right., \label{sol-in}
\end{align}
Near the event horizon there is no outgoing wave whereas at the cosmic horizon there are both incoming and outgoing modes.

When $\Lambda>0$, the background metric $f(r)$ has four possible real roots namely, $r_{-}$, $r_{C}$, $r_{H}$, $r_{\Lambda}$. As shown in Section \ref{setup}, there are two kinds of extremality for this background; $r_{C}\sim r_{H}$ and $r_{H}\sim r_{\Lambda}$. For the case of $r_{C}\sim r_{H}$, $A$ and $B$ are chosen to be $r_H$ and $r_C$ respectively while $C$ and $D$ can be either $r_{-}$ or $r_{\Lambda}$. Since the physical Universe is covered by only one extremal horizon, the term $(r-r_{-})(r-r_{\Lambda})$ must be expanded around $r_{H}$. The resultant quasinormal frequencies are thus given by (\ref{FM}),
	\begin{equation}\label{w1}
		\omega_{n}=\kappa_{H}\left\lbrace\sqrt{\frac{V_{0}}{\kappa^{2}_{H}}-\frac{1}{4}}-\left(n+\frac{1}{2}\right)i\right\rbrace,
	\end{equation}
where,
	\begin{equation}\label{Vw1}
		V_{0}=\frac{\kappa^{2}_{H}}{(r_{H}\gamma+\tau-2Q^{2}/r^{2}_{H})}\Big[m^{2}_{s}r^{2}_{H}+l(l+1)\Big].
	\end{equation}

As for the case of $r_{H}\sim r_{\Lambda}$, we choose $A$ to be $r_H$ and $B$ to be $r_\Lambda$. $C$ and $D$ are mapped to either $r_{-}$ or $r_{C}$. With the physical universe enveloped by extremal horizons, the tortoise coordinate can be directly calculated from $\displaystyle{dr_{*}=\frac{dr}{f(r)}}$. The quasinormal frequencies are given by exactly the same formulae (\ref{w1}) and (\ref{Vw1}).

\subsection{Solutions in Asymptotically AdS Background}

For this case, the boundary condition at the event horizon is the incoming waves and at spatial infinity is zero for the scenario where the physical universe is in $r>r_{\Lambda+}$ region. However, the boundary condition changes to those of the asymptotically dS space given by (\ref{sol-in}) when we consider the scenario where universe has two horizons, i.e., in the region $r_{H}<r<r_{\Lambda-}$, with $r_{\Lambda-}$ playing the role of cosmic horizon.

The metric function $f(r)$ can possess up to four distinct horizons when $\Lambda<0$ i.e., $r_{C}$, $r_{H}$, $r_{\Lambda-}$ and $r_{\Lambda+}$. $r_{C}$ and $r_{H}$ behave identically to their counterpart in the dS background. $r_{\Lambda-}$ is akin to $r_{\Lambda}$ from the dS background. $r_{\Lambda+}$ is event horizon for the unbounded physical universe. While there are varieties of near-extremal black holes shown in Section \ref{setup}, for the purpose of finding the QNM formula; there are only three kinds of near-extremality we will consider. They are $r_{C}\sim r_{H}$, $r_{H}\sim r_{\Lambda-}$ and $r_{\Lambda-}\sim r_{\Lambda+}$ scenarios. 

We now consider the near-extremal case where $r_{C}\sim r_{H}$ and $r_{H}\sim r_{\Lambda-}$. For the latter case, the parameter $A$ is set to $r_{H}$ and $B$ is set to $r_{\Lambda-}$. It turns out that the quasinormal frequencies formulas for both cases are identical to those of the dS case (\ref{w1}).

For the final case where $r_{\Lambda-}\sim r_{\Lambda+}$, the root parameter $A$ is set to be $r_{\Lambda+}$ and $B$ is set to be $r_{\Lambda-}$. The quasinormal frequency is,
	\begin{equation}
		\omega_{n}=\kappa_{\Lambda+}\left\lbrace\sqrt{\frac{V_{0}}{\kappa^{2}_{\Lambda+}}-\frac{1}{4}}-\left(n+\frac{1}{2}\right)i\right\rbrace,\nonumber
	\end{equation}
where
	\begin{equation}
		V_{0}=\frac{\kappa^{2}_{\Lambda+}}{(r_{\Lambda+}\gamma+\tau-2Q^{2}/r^{2}_{\Lambda+})}\Big[m^{2}_{s}r^{2}_{\Lambda+}+l(l+1)\Big].\label{FMsp}
	\end{equation}
There is one special characteristic of quasinormal mode in this particular case. The pair of near-extremal horizons is connected to two different physical Universes. Since there is no physical constraints to prevent the above formula from functioning in both physical universes, the above formula is valid for QNMs in both regions. This statement leads to a very interesting conclusion. The inner physical Universe cannot perceive the existence of the $r_{\Lambda+}$ and the external ``universe''. However, the quasinormal modes of the cosmic horizon in this Universe~($r_{\Lambda-}$) depends on the surface gravity of the associated horizon $r_{\Lambda+}\simeq r_{\Lambda-}$ in such a way that it can be determined very accurately by (\ref{FMsp}), giving way for internal observers to {\it know} the extremality of their cosmic horizon with respect to the external universe in $r>r_{\Lambda+}$ region.  In this sense even though covered by horizons, two physical spacetime regions or ``universes'' are globally connected via the unique QNMs.

\section{Analysis of Strong Cosmic Censorship Conjecture}  \label{SCCSec}

Existence of singularity at $r=0$ in the metric of a spherically symmetric black hole spacetime predicts its own domain of validity of the underlying gravity theory at the classical level.  The existence of singularity would represent the break down of causality and unitarity as well as the absurd infinite density. Conventional point of view is the rescue~(or worsening) of quantum gravity around the Planck scale where the singularity would be replaced by some unknown states. 

Classically, the problem of singularity can be evaded by a number of proposals. For example, wormhole connecting to other spacetime region or other universe can be patched to the black hole solution as the inner region of the horizon, replacing the problematic region of spacetime containing the singularity. Bag of gold solutions~(or monsters~\cite{Ong:2013mba}) patching to the horizon resolve the singularity with bag of spacetime with arbitrary size and entropy.  These classically patched solutions could possibly be thought of as the evolved states of energy-matter and spacetime after they collapsed to form black hole in the outer region and reached highly densed quantum gravity states in the inner region of the horizon.  In terms of the original metric solution containing a classical singularity, only the outside metric is physically unchanged~(modulo the tiny quantum effect of Hawking radiation) while the unknown situations inside are irrelevant as long as there is a horizon dividing the outside classical reliable solution and the inside unknown replacement of the classical singularity.  Cosmic Censorship Conjecture was proposed to ensure the classical determinism of GR and guarantee independence of physics outside the horizon from the inside. Specifically, SCC states that the maximal Cauchy development of initial data is inextendible and classical determinism of General Relativity is preserved. On the other hand, WCC states that if there exist singularities, they must be hidden behind horizons and cannot be extended to the future null infinity.

For a charged black hole or Reissner-Nordstrom solution in GR, the central singularity is always covered by a Cauchy horizon for sub-extremal black holes.  This Cauchy horizon naturally blueshifts any finite incoming waves from the outer region to an unbound magnitude rendering the background spacetime unstable.  The instability leads to inextendibility of spacetime beyond the Cauchy horizon and thus preserves the SCC.  As such, any singularity of the charged black hole should be covered by an unstable Cauchy horizon. However, the notion of ``inextendibility'' of spacetime has some subtlety, e.g. for smooth initial data with $1>\beta >1/2$~(see definition below), the Cauchy horizon could be $C^{r}~(r\geq 1)$ inextendible while allowing the finite energy solution to pass across~\cite{Dias:2018etb}. Also, the SCC could be respected for non-smooth initial data while violated for the smooth ones~\cite{Dafermos:2018tha}.

Following a proposal in Ref.~\cite{Cardoso:2017soq,Hintz:2015jkj}, the validity of $C^{r}$ and Christodoulou~\cite{Christodoulou:2008nj} versions of SCC for smooth initial data is determined by a parameter
	\begin{equation}
		\beta=\frac{\alpha}{|\kappa_{C}|},
	\end{equation}
where the {\it spectral gap} $\alpha =-{\rm Im}~\omega$ of the longest living quasinormal mode and $\kappa_{C}$ is the surface gravity of Cauchy horizon. With $\beta <\frac{1}{2}$, the Christodoulou version of SCC~(CSCC) holds. While the $C^{1}$-SCC version only requires that $\beta <1$. From our formula for quasinormal mode frequencies the parameter $\alpha$ is,
	\begin{equation}
		\alpha=\text{inf}\left\lbrace-\kappa_{\chi}\text{Im}\left[\pm\sqrt{\frac{V_{0}(m_{s},l)}{\kappa^{2}_{\chi}}-\frac{1}{4}}-\left(n+\frac{1}{2}\right)i\right]\right\rbrace,
	\end{equation}
where $\kappa_{\chi}$ is the surface gravity of the near-extremal horizon. The symbol inf is called infimum that takes the smallest value of a subset. For the purpose of this paper, this symbol represents the smallest value of $\displaystyle{-\text{Im}~\omega_{n,l}}$ provided that the value is still positive. If $-\text{Im}~\omega_{n,l}$ is zero or negative, then the quasinormal mode is a constant or growing mode, which violates our definition of $\alpha$.

This $\beta$-criteria is based on the consideration of RN-dS black hole in GR, a rigorous proof is given in Ref.~\cite{Hintz:2015jkj}. The assumption of the proof requires essentially only the existence of 3 positive real roots of the metric function $f(r)$, the Cauchy horizon $r_{C}$, the event horizon $r_{H}$ and the cosmic horizon $r_{\Lambda}$ with condition $f(r)>0$ in $r<r_{C}, r_{H}<r<r_{\Lambda}$ and $f(r)<0$ in $r_{C}<r<r_{H}$ region.  The RN-dS metric function in GR actually can be rewritten in the form of generalized metric function given by (\ref{gmet1}). In the range of parameters suitable for SCC consideration, this generalized metric function (\ref{gmet1})~(for positive $\Lambda$) has 3 positive real roots and 1 negative root~(which is unphysical and irrelevant in the proof given in \cite{Hintz:2015jkj}). It satisfies the underlying assumption of the proof.  The metric function of RN-dS in GR takes exactly the same form as (\ref{gmet1}) with one extra condition, $A+B+C+D=0$. This extra condition is not required by any argument given in the proof of Ref.~\cite{Hintz:2015jkj} as long as the metric function (\ref{gmet1}) has 3 positive real roots satisfying the above conditions.  Moreover, the equation of motion (\ref{KGeq}) of the scalar field in the generalized background is also identical to the GR case with only difference in the metric function. Consequently, all the arguments in the original GR proof should be applicable to the form of metric function (\ref{gmet1}), and exactly the same $\beta$-criteria should also be valid for the scalar perturbation in the generalized metric in our case. 

Another argument to support the validity of the above $\beta$-criteria for the generalized background given by the metric function (\ref{gmet1}) is based on the equation of motion (\ref{reom}). The equation takes the exact form with the P\"{o}schl-Teller potential and has exact solution in the same form as the corresponding solution in the GR background. Consequently, the solutions must satisfy the same $\beta$-criteria as in the near-extremal GR solution.

It is shown in the master equation for the near-extremal quasinormal mode~(\ref{FM}) that $\omega_{n}$ depends on $\kappa_{\chi}$, where $\kappa_{\chi}$ is the surface gravity of a near-extremal horizon whose value is positive. If the pair of extremal horizons does not contain a Cauchy horizon, then the following property holds, 
	\begin{equation}
		\beta\sim\frac{\kappa_{\chi}}{|\kappa_{C}|}\ll 1,\label{smUc}
	\end{equation}
since the surface gravity of near-extremal horizons are much less than the surface gravity of the Cauchy horizon. Therefore, both CSCC and $C^{1}$-SCC hold for all near-extremal black holes whose Cauchy horizon is not a near-extremal horizon. Hence, for the rest of this Section, only the case where Cauchy horizon is also a near-extremal horizon is considered.

From Section \ref{setup}, multitude of loci of extremal condition are demonstrated in Fig.~\ref{QGPSdS} and \ref{QGPSAdS} for the dS and AdS cases respectively. By shifting slightly from these locus lines, loci of near-extremal condition are obtained. The value of $\beta$ can be evaluated along these loci and both conjectures can be subsequently verified.  If the term under the square root sign of the QNM expression is positive, the formula of $\beta$ is then simplified as
	\begin{equation}
		\beta=\frac{\kappa_{\chi}}{|\kappa_{C}|}\left(n+\frac{1}{2}\right).\nonumber
	\end{equation}
It is clear that $n$ must be set to zero. This setting leads to the following relation, $\beta\lesssim 1/2$, implying that both CSCC and $C^{1}$-SCC are not violated. Fortunately for $\kappa_{C}\notin\text{Near-Extremal horizon}$, all kinds of near-extremal background satisfy the following conditions simultaneously,
	\begin{eqnarray}
		\frac{V_{0}(l)}{\kappa^{2}_{\chi}}-\frac{1}{4}&>&0,\nonumber 
	\end{eqnarray}
and thus CSCC and $C^{1}$-SCC hold.

For the case where the term under the square root of Eqn.~(\ref{FM}) is negative, the value of $\beta$ is much more complicated to inspect. Let us choose the negative sign for the square root sign~(i.e. the square root gives negative imaginary value). For this case, the $\beta$ has the minimum value if $n=0$. Since both $\kappa_{\chi}$ and $\kappa_{C}$ are both extremal horizons, $\kappa_{\chi}/\kappa_{C}\lesssim1$. While the exact value of the term under the square root is yet to be calculated, we know from numerical calculation that for near-extremal black holes, $V_{0}/\kappa^{2}_{\chi}<0$~(see Fig. \ref{HdS} and \ref{HAdS} where $-H\propto V_{0}/\kappa^{2}_{\chi}$ with positive proportionality constant) and thus $\beta\ge1$ if the negative sign of the square root is taken. Hence, the positive sign of the square root must be chosen to explore the validity of both versions of SCC. Since the mass of black hole $M$ is set to one and the ratio of scalar mass to the black hole mass should be small to avoid the backreaction, $m_{s}\ll 1$ approximation can be imposed.  Moreover, because only the $m^{2}_{s}\ll 1$ term appears in the formula for $\beta$, the massless approximation can be employed without much loss of generality. With the relation, $R\equiv \kappa_{\chi}/|\kappa_{C}|\lesssim1$~(can be made arbitrarily close to 1 as we approach extremality), the formula for $\beta$ is simplified as the following,
	\begin{equation}
		\frac{|\kappa_{C}|}{\kappa_{\chi}}\beta=\left(n+\frac{1}{2}\right)-\sqrt{\frac{1}{4}-\frac{l(l+1)}{r_{\chi}\gamma+\tau-2Q^{2}/r^{2}_{\chi}}}.\nonumber
	\end{equation}
We then define an auxiliary function $\displaystyle{H(\tau,M,Q,\gamma,\Lambda)=-\frac{1}{r_{\chi}\gamma+\tau-2Q^{2}/r^{2}_{\chi}}}$. This function contains all the spacetime parameters. The $\beta$ becomes,
	\begin{equation}
		\frac{|\kappa_{C}|}{\kappa_{\chi}}\beta=\left(n+\frac{1}{2}\right)-\sqrt{\frac{1}{4}+Hl(l+1)}.
	\end{equation}
From the condition of CSCC,
\be
		0<\frac{|\kappa_{C}|}{\kappa_{\chi}}\beta<\frac{1}{2R},\nonumber 
\ee
leading to the constraint
\be		
		\frac{n(n+1)}{l(l+1)}>H>\frac{n(n+1)}{l(l+1)}-\frac{\left( n+\frac{1}{2}-\frac{1}{4R}\right)}{R~l(l+1)}.\label{CS}
\ee
The parameter $H$ is then being calculated along the near-extremal loci given in Fig.~\ref{QGPSdS} and \ref{QGPSAdS}. By simply show that the upper and lower bounds from (\ref{CS}) can cover all ranges of possible $H$, the CSCC would be proven to be valid. For later use, we define the LHS~(RHS) of inequality (\ref{CS}) as UB~(LB) respectively. Since $R$ can be set arbitrarily close to 1 as we approach extremality $r_{H}\to r_{C}$, first we will set $R=1$ in the calculation of the bounds in Sect.~\ref{SCCdS}-\ref{SCCGR}. The resulting bounds will be applicable to extremal BHs but need modifications for near-extremals. Typically for $r_{H}-r_{C}\sim 10^{-2}, R\sim 0.93$ resulting in $1\%$ decrease in the lower bounds and certain gaps in the bounds will be closed as will be shown in Section \ref{RSec}. Appendix~\ref{Rproof} proves that $R<1$ for $r_{H}\simeq r_{C}$ near-extremals in the presence of the cosmic horizon which is the only relevant case for SCC consideration.

To prove that CSCC is {\it protected}, we need to find any combinations of $(n,l)$ each of which give the upper and lower bounds, UB and LB, that cover all possible values of $H$. If there is a gap in $H$ value that cannot be covered by any $(n,l)$, then the CSCC is violated for the BH parameters that give that particular gap value of $H$.  First, in each of the asymptotically dS, AdS, and $\gamma = 0$~(GR) cases in Sect.~\ref{SCCdS}, \ref{SCCASec}, \ref{SCCGR}, we consider CSCC of extremal BHs by assuming $R=1$. Later in Sect.~\ref{RSec}, the non-extremal effect $R<1$ will be taken into account. It will be shown that certain extremal gaps where CSCC is violated are closed up. However, CSCC violations still exist for the remaining gaps in the near-extremal cases. On the other hand, $C^{1}$-SCC is always valid since $\beta <1$ in all cases.

\subsection{The Strong Cosmic Censorship Conjectures for near-extremal Black Holes in asymptotically dS space}  \label{SCCdS}

In this case, only the near-extremal black holes of the kind $r_{C}\sim r_{H}$ need to be considered. The CSCC for $r_{H}\sim r_{\Lambda}$ near-extremal case automatically holds since it satisfies (\ref{smUc}). By setting $M=1$ and $\Lambda=0.03$, the extremal line can be plotted in $Q^{2}$-$\gamma$ parameter space as shown in Fig. \ref{QGPSdS}. The parameter $H$ is then being calculated along the near-extremal locus in Fig. \ref{QGPSdS} (a) and (b). The values of $H$ of extremal BHs are covered by the range of $H$ obtained along the near-extremal locii. 

\begin{figure}[h]
		\centering
		\includegraphics[width=0.7\textwidth]{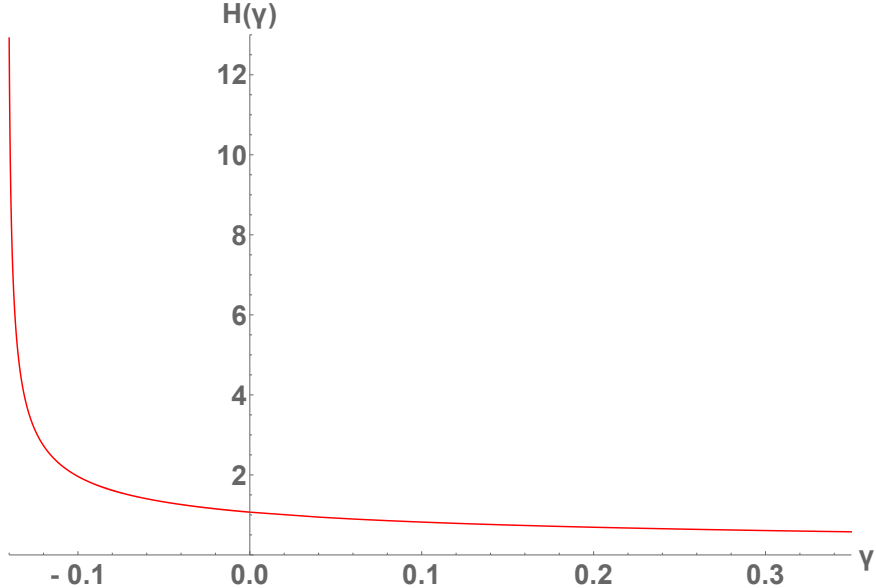}
		\caption{An example of $H$ calculated along near-extremal line $r_{C}\sim r_{H}$. The calculation is performed with $M=1$ and $\Lambda=0.03$.}
		\label{HdS}
	\end{figure}

The plot between $H$ and $\gamma$ is shown in Fig. \ref{HdS}. The value of $H$ monotonically increases with the negative $\gamma$, the more negative $\gamma$, the larger value of $H$. $H$ increases without bound as the end of extremal line $r_{H}=r_{C}$ is approached. For $\gamma \geq 0$, the sets of $n$ and $l$ shown in Table \ref{BETAdS} are used to calculate the upper and lower bounds of $H$ that cover all of the range $\frac{35}{288}\simeq 0.1215 < H < 1$~(the lower bound of one set is less than the upper bound of the next, i.e., they overlap, therefore all sets cover all of the range of $H$ in this region). 
	\begin{table}[h]
		\centering\
		\begin{tabular}{|c|c|c|}
			\hline
			~$(n,l)$~~&$\frac{n^{2}-1/4}{l(l+1)}$~~&$\frac{n(n+1)}{l(l+1)}$~~\\
			\hline
			(1,1)&$\frac{3}{8}$&$1$\\
			\hline
			(3,5)&$\frac{7}{24}$&$\frac{2}{5}$\\
			\hline
			(1,2)&$\frac{1}{8}$&$\frac{1}{3}$\\
			\hline
			(3,8)&$\frac{35}{288}\simeq 0.1215$&$\frac{1}{6}$\\
			\hline
		\end{tabular}
		\caption{Upper and lower bounds of $H$ covering $\frac{35}{288}\simeq 0.1215 < H < 1$ region.}
		\label{BETAdS}
	\end{table}
Similarly various sets of $(n,l)$ are needed to cover the range of $H$ for $\gamma < 0$ region.  More robust investigation using code to generate UB and LB to cover all possible gaps are shown in Fig.~\ref{boundsfig} of Appendix~\ref{SCCP}. The plot reveals four gaps around $H=1, 4, 16, 36$. These gaps can eventually be closed up to exactly only 4 values, $H=1,4,16,36$ in the extremal limits as explained below.

For $n=l+c$ for some constant $c\in \mathbb{Z}$, both the lower and upper bounds, $\displaystyle{\frac{n^{2}-1/4}{l(l+1)}}$ and $\displaystyle{\frac{n(n+1)}{l(l+1)}}$ in the extremal limits, can be made to approach $1$ as close as desired in the large $(n,l)$ limit, see the proof in Appendix \ref{SCCP}. For example for $(n,l)=(200000,199999)$, the lower and upper bounds are 1.000005 and 1.00001 respectively. On the other hand for $(n,l)=(200000,200000)$, the lower and upper bounds are 1 and 0.999995. Curiously, the {\it open} bounds never cover $H=1$

Similarly, other gaps around $H=4,16,36$ can be closed up until only the points $H=4,16,36$ remain not being covered by the bounds. From Appendix \ref{SCCP} in the extremal limits, we see that for the relevant region under consideration e.g. $H<40$, only the values $H=1,4,16,36$ are not covered by any sets of $(n,l)$. At these points,{\it the CSCC is violated}. On the other hand, there is one set of $(n,l)=(0,0)$~(after retrieving the mass term $m_{s}^{2}$ in $V_{0}$) that protects $C^{1}$-SCC for any values of $\gamma$. However, the gaps at $H=4, 16 , 36$ will be shown to close up for near-extremals in Sect.~\ref{RSec}.

\subsection{The Strong Cosmic Censorship Conjectures for Black Hole in Anti de-Sitter space}  \label{SCCASec}

The proof of the $\beta$-criteria for the SCC is based on the asymptotically dS space. However in the AdS scenario where there exist 4 positive real roots, the SCC analysis can still be applied to the Cauchy horizon since the spacetime beyond $r_{\Lambda_{-}}$ is not causally connected to the Cauchy horizon and no signals from $r>r_{\Lambda_{-}}$ can be transmitted to $r_{C}$ and destabilize it.  The spacetime involved in the stability of Cauchy horizon is effectively an asymptotically dS space in $0<r<r_{\Lambda_{-}}$ region. We can thus apply the $\beta$-criteria for SCC in such scenario.  

For the asymptotically Anti de-Sitter background, it is best to differentiate a case where $\gamma$ is positive and negative. If $\gamma$ is positive, the only term in the metric $f(r)$ whose value can be negative is $-2M/r$ term. While this suffices to create a region with negative value of $f(r)$ and two horizons, it is not enough to allow existence of more than two positive real roots. Hence, there is only one case of near-extremal black hole to be considered, $r_{C}\sim r_{H}$. As shown in Fig.~\ref{QGPSdS} and \ref{QGPSAdS}, the near-extremal conditions of this kind are each represented by a curve that is quite similar to the asymptotically de-Sitter case. Since the numerical value of $\Lambda$ is small, the effect of $\Lambda$ is dominated by all other parameters. Hence, the value of $H$ along this line behaves very similar to its counterpart in the de-Sitter case. The sets of bounds from the previous section can be used to verify the validity of the SCCs for positive $\gamma$ case.

	\begin{figure}[h]
		\centering
		\includegraphics[width=0.7\textwidth]{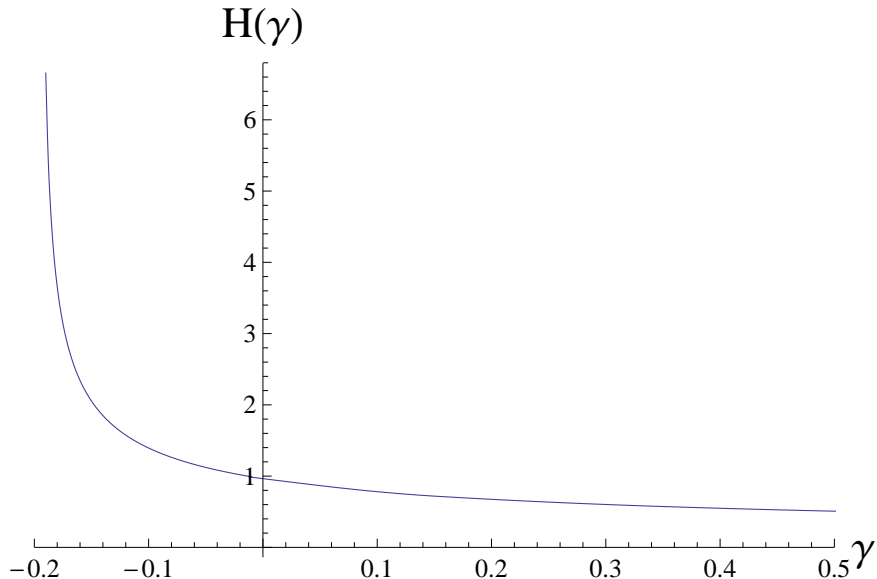}
		\caption{An example of $H$ calculated along near-extremal line $r_{C}\sim r_{H}$. The calculation is performed with $M=1$ and $\Lambda= -0.03$.}
		\label{HAdS}
	\end{figure}

For $\gamma<0$, there are two curves of near-extremality to be inspected. As shown in Fig. \ref{QGPSAdS}, there are three loci of extremality. However, two extremal lines on the left and right sides of the ``All Four Real Roots'' region embody the extremal of the type $r_{H} = r_{\Lambda-}$ and $r_{C}<r_{\Lambda-}= r_{\Lambda+}$, the extremal conditions which automatically satisfy the SCCs since Cauchy horizon is not one of the extremal horizons, i.e., (\ref{smUc}). There are then two remaining extremal lines to be considered, the left-to-right line and the vertical line starting from the double-extremal point upward~(where $r_{C}=r_{\Lambda-}<r_{H}=r_{\Lambda+}$ on the left side of the line). 

First, we evaluate parameter $H$ along the left-to-right extremal curve of Fig.\ref{QGPSAdS}~(b). The near-extremal condition is $r_{C}\sim r_{H}$ along this line. The value of $\gamma$ is then changed within the range $\gamma >-0.19462995$. If the value of $\gamma$ falls below $-0.19462995$, $r_{H}$ would cease to exist. Along the locus of near-extremal condition, the value of $H$ ranges from $0.9644$~(at $\gamma =0$) to arbitrarily large value as $\gamma \to-0.19462995$. The value of $H$ along this extremal condition is similar to its counterpart in de-Sitter case as shown in Fig.~\ref{HAdS}. Hence, the sets of $(n,l)$ from the dS case can be used to cover most of the range of $H$ here except at $H=4,16,36$. By using all possible bounds within $n=1-150,l=1-150$ as shown in Fig.~\ref{boundsfig}, most of the gaps will be closed up except around certain values. Similar to the dS case, there are exactly the same points $H=1,4,16,36$~(in the extremal limits) where CSCC are violated. The $C^{1}$-SCC is protected by the same reason, $(0,0)$~(after retrieving the mass term $m_{s}^{2}\ll 1$ in $V_{0}$) validates the bounds (\ref{CS}).

Lastly, the vertical extremal line in Fig. \ref{QGPSAdS}~(b) from the double-extremal point upward is considered. The domain of $\gamma$ is $\gamma < -0.17577701335$ for this case. Function $H$ still has a similar monotonic property in the range $H<3.151425$ where $H=3.151425$ at the double-extremal point. 
Again, this range of H can be closed up by appropriate choices of sufficiently large $(n,l)$, see Appendix \ref{SCCP}. The only remaining point not covered in this interval is $H=1$, see the corresponding extremal black hole parameters in Table~\ref{sumtab1}. The extremal BHs at $H=1$, however, have no cosmic horizon and are not subject to the $\beta$-criteria and whether CSCC is violated or not is still an open question. Mass inflation similar to the asymptotically flat RN BH discussed in Ref.~\cite{Luk:2015qja} might be able to protect or still violate the CSCC if the resulting singularity at the Cauchy horizon is weak~\cite{Bhattacharjee:2016zof}. On the other hand, the extremals at $H=4,16,36$ have cosmic horizon $r_{\Lambda-}$~(in addition to $r_{C}$ and $r_{H}$) and the CSCC is violated according to the $\beta$-criteria. In summary for AdS case, the CSCC is potentially violated at $H=1,4,16,36$ and probably continuing further while the $C^{1}$-SCC is still valid as in the dS case. For near-extremals however, the gaps at $H=4, 16, 36$ will be closed up as discussed in Sect.~\ref{RSec}.

\subsection{SCC for $\gamma=0$ spacetime, the charged black hole in GR case}  \label{SCCGR}

In this section the value of $H$ is being evaluated when $\gamma$ is set to zero. By setting the parameter $\gamma$ to zero, the obtained numerical parameters are functions of only $M$, $Q$ and $\Lambda$. With the value of $M$ set to unity, the near-extremal condition acts as a constraint relating $Q$ to $\Lambda$. We are only interested in the near-extremal black holes with $r_{C}\sim r_{H}$. For $-0.05<\Lambda <0.05$, the range of $H$ is $0.932375<H<1.14382$ for $r_{H}-r_{C}=10^{-2}$. According to our previous investigation into the bounds of $H$ which depend only on $n$ and $l$, the only possibilities for the violation of CSCC are when $H=1,4,16,36$, see Appendix \ref{SCCP}. As shown in Fig.~\ref{SCCzerogam}, the corresponding black hole parameters at $H=1$ are $M=1$, $\gamma=0$, $Q^{2}=0.996602$ and $\Lambda=-0.010255$ for near-extremal black hole with $r_{H}-r_{C}=10^{-2}$. This is a mere example amongst infinitely many other possibilities of near-extremal black holes that potentially violates CSCC depending on the values of $\epsilon=r_{H}-r_{C}$ for AdS case~($\Lambda <0$ side of Fig.~\ref{SCCzerogam}). These BHs, however, do not have cosmic horizon and mass inflation discussed in Ref.~\cite{Luk:2015qja} again might be able to protect or still violate the CSCC if the resulting singularity at the Cauchy horizon is weak~\cite{Bhattacharjee:2016zof}. It is still an open question whether CSCC is violated in this case. 

	\begin{figure}[h]
		\centering
		\includegraphics[width=0.8\textwidth]{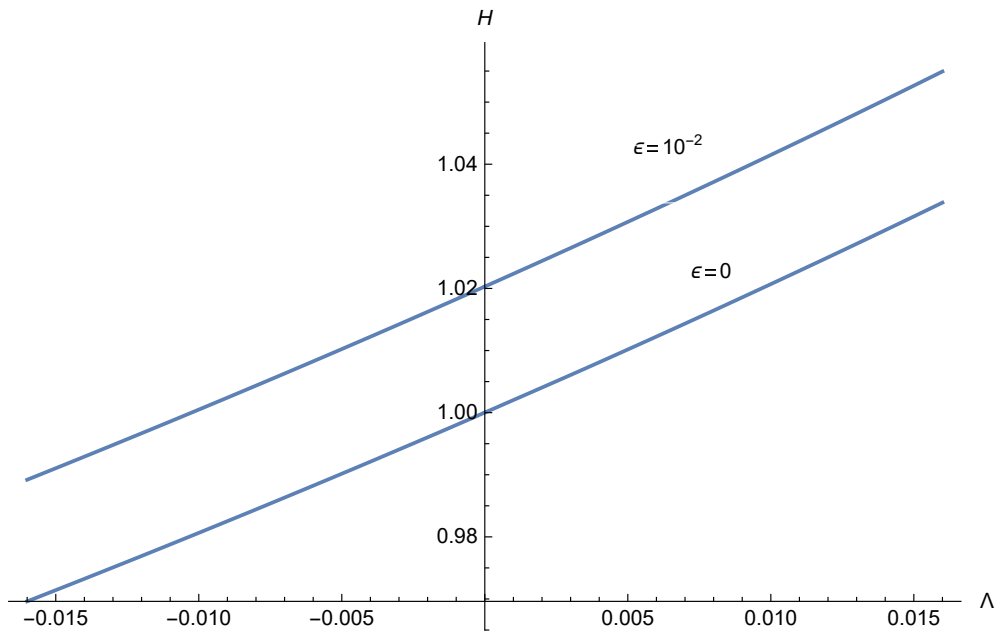}
		\caption{The value of $H$ calculated along near-extremal line $r_{H}-r_{C}=\epsilon$ for $M=1$ and $\gamma=0$. For $\epsilon =0$, it is the extremal RN black hole in GR with $H=1$ at $Q=1$.}  \label{SCCzerogam}
	\end{figure}

	\begin{table}[h]
		\centering\
		\begin{tabular}{|c|c|c|}
			\hline
			~Spacetime/BH type~~&~Cauchy $\sim ~(=)$ Event horizon~&~Event $\sim~(=)$ Cosmic horizon~\\
			\hline
			$\gamma \neq 0$ dS& CSCC violated at $H=1~(,4,16,36)$&$\beta \ll 1$ CSCC valid\\
			\hline
			$\gamma \neq 0$ AdS&CSCC violated at $H=1~(,4,16,36)$&$\beta \ll 1$ CSCC valid\\
			\hline
			$\gamma=0$&CSCC violated at $H=(4,16,36)$&N/A for $\Lambda M^{2}<\frac{1}{9}$~(Narai limit at $Q=0$)\\
			& & CSCC valid for $\Lambda M^{2}=\frac{2}{9}, \frac{Q^{2}}{M^{2}}=\frac{9}{8}$ extremal\\
			\hline
		\end{tabular}
		\caption{Summary Table of potential SCC violations, $C^{1}$-SCC is always valid.}
		\label{sumtab}
	\end{table}
For dS~($\Lambda >0$ side of Fig.~\ref{SCCzerogam}), along the extremal contour we found that $H$ can go arbitrarily large~(numerically verified up to $H=10^{5}$) covering all $H=4,16,36$ points that CSCC is violated. Therefore we have found a series of extremal {\it RN-dS black holes in GR} corresponding to $H=4,16,36$~(probably continuing indefinitely) that violate CSCC. However, in Sect.~\ref{RSec} for near-extremals with $R<1$, these $H=4,16,36$ point-wise gaps will be completely closed due to decrease of the LBs. At exactly $\Lambda = 0$, the gap is at $H=1$, the conventional extremal RN BH. Mass inflation should protect the CSCC in this case~\cite{Luk:2015qja}.

	\begin{table}[h]
		\centering\
		\begin{tabular}{|c|c|c|c|}
			\hline
			~$H$~~&~$\Lambda= -0.03$~&~$\Lambda= 0.03$~&~$\gamma =0$~(GR)\\
			\hline
			$1$ & $Q^{2}=1.010313350629,\gamma=-0.020207$ & $Q^{2}=0.9902891778363,$& $\Lambda=0,Q^{2}=1$\\
			& $Q^{2}=9.45644641063,\gamma=-0.1891298$ &$\gamma=0.019805097276985394$ & \\
			& (no cosmic horizon) &&(no cosmic horizon)\\
			\hline
			$4$ & $Q^{2}=1.324425000,$ & $Q^{2}=1.244413039569,$ & $\Lambda= 0.2109375,$ \\
			&$\gamma=-0.1827385$&$\gamma=-0.13136172$& $Q^{2}=1.111111111111111$ \\		
			\hline
			$16$ &$Q^{2}=1.3907719269,$ & $Q^{2}=1.28350241682389,$ & $\Lambda=0.221557615,$\\
			&$\gamma=-0.19383106$&$\gamma= -0.14034556589$&$Q^{2}=1.1239669388859$ \\
			\hline
			$36$&$Q^{2}=1.3965335598307,$ & $Q^{2}=1.286225225062,$ & $\Lambda=0.222092244789899291,$\\
			& $\gamma=-0.19446871991856804$ &$\gamma=-0.1408134998$&$Q^{2}=1.12478859417527$ \\
			\hline
		\end{tabular}
		\caption{Summary Table of {\it extremal} BH parameters with potential CSCC violation for $M=1, r_{H}=r_{C}$, $C^{1}$-SCC is always valid. Extremal AdS BHs and extremal RN BH at $H=1$ have no cosmic horizon and thus are not subject to the $\beta$ criteria. CSCC should be protected by mass inflation~\cite{Luk:2015qja} for the extremal RN BH case. }
		\label{sumtab1}
	\end{table}

\subsection{Effects of $R$ on SCC gaps}  \label{RSec}

For near-extremal BHs with small $\epsilon =r_{H}-r_{C}\ll 1$ and $R<1$, the second term of the LB~(now defined as the RHS of (\ref{CS})) could become so large that the UB~(independent of $R$) and LB would cover each of the extremal gaps $H=4,16,36$ in contrast to the extremal bounds in Appendix~\ref{SCCP}. To see this, lets rewrite $n=kl+c$, for $k=2,4,6$ let $c=k/2$ and consider separation between LB and its limit $k^{2}$ in the large $l$ and $a\equiv 1-R \ll 1$ limit,
\bea
{\rm LB}-k^{2}&=&\frac{n(n+1)}{l(l+1)}-\frac{\Big( n+\frac{1}{2}-\frac{1}{4R}\Big)}{R~l(l+1)}-k^{2}  \notag \\
			    &=&\frac{k^{2}-2ka(2 l+1) -1}{4l(l+1)}+\mathcal{O}(a^{2}). \label{rgap}
\eea
For large $l$, the quantity on the RHS of (\ref{rgap}) could be negative for $l>(k^{2}-1)/4ka$ which means that LB is lower than $k^{2}$ while UB$>k^{2}$, covering the gap at UB$>H=k^{2}>$LB. Specifically for $R=0.93$, $k=2,4,6$, $c=1,2,3$, and $l=100$, LB$=3.99444,~15.9942,~35.9988$ while UB$>4, 16, 36$ respectively. Namely the gaps at $H=4, 16, 36$ are closed for $R<1$ for sufficiently large $l$.  

On the other hand for $n=l+c,$ we know from Appendix~\ref{SCCP} that the only possibility to close the gap is when $c\geq 1$ so that the LB drops below 1 while keeping UB$>1$. Again, consider separation between LB and its limit $k^{2}=1$ in the large $l$ and $a\equiv 1-R \ll 1$ limit,
\bea
{\rm LB}-1^{2}&=&\frac{n(n+1)}{l(l+1)}-\frac{\Big( n+\frac{1}{2}-\frac{1}{4R}\Big)}{R~l(l+1)}-1^{2}  \notag \\
			    &=&-\frac{1-4 c^{2}+4 l(1-2c+ a)+4 ca}{4 l(l+1)}+\mathcal{O}(a^{2}). \label{rgap1}
\eea
The RHS of (\ref{rgap1}) needs to be negative for $c\geq 1$ but this implies that
\be
l < \frac{4c^{2}-1-4ca}{4-8c+4a},
\ee
which is impossible since $l>0$~(the numerator~(denominator) is always positive~(negative)). The LB thus cannot drop below 1 even for $R<1$. This proves that even for near-extremal with $R<1, a\ll 1$, the point-wise gap at $H=1$ cannot be closed.  
	
		\begin{table}[h]
		\centering\
		\begin{tabular}{|c|c|c|c|}
			\hline
			~$H$~~&~$\Lambda= -0.03$~&~$\Lambda= 0.03$~&~$\gamma =0$~(GR)\\
			\hline
			$1$ & $Q^{2}=1.0033445,\gamma=-0.01337$ & $Q^{2}=0.983654,\gamma=0.026715$& $\Lambda=-0.010255,Q^{2}=0.996602$\\
			& $Q^{2}=9.44365626,\gamma=-0.1891147$ &&\\
			\hline
		\end{tabular}
		\caption{Summary Table of {\it near-extremal} BH parameters with potential CSCC violation for $M=1, r_{H}-r_{C}=10^{-2}$, $C^{1}$-SCC is always valid. BHs with $\Lambda<0$ in this Table have no cosmic horizon and are not subject to the $\beta$-criteria. }
		\label{sumtab2}
	\end{table}
	
Violations of CSCC are summarized in Table~\ref{sumtab},\ref{sumtab1} and \ref{sumtab2}.

\section{Black Strings in the Near-Extremal limits}  \label{bstring}

In higher dimensions, a black string in the braneworld scenario can exist as a result of gravitational collapse occured in the brane.  The black string often suffers from Gregory-Laflamme instability which can be studied via its QNMs~\cite{Konoplya:2008yy}. In standard general relativity~(GR) with negative cosmological constant, black string can exist as a cylindrically symmetric solution. However, in the dRGT massive gravity theory, the effect of graviton mass also allows the existence of black string in the asymptotically dS background with positive cosmological constant which is not possible in GR. The metric of the dRGT black string is given by \cite{Tannukij:2017jtn},
	\begin{equation}
		ds^{2}=-f(r)dt^{2}+\frac{dr^{2}}{f(r)}+r^{2}\big(\alpha^{2}_{g}dz^{2}+d\varphi^{2}\big),
	\end{equation}
where
	\begin{equation}
		f(r)=\alpha^{2}_{m}r^{2}-\frac{2M}{r}+\frac{Q^{2}}{r^{2}}+\gamma r+\epsilon_{0}.
	\end{equation}
The Klein-Gordon equation in radial direction takes the form~ \cite{Ponglertsakul:2018smo},
	\begin{equation}
		\frac{d^{2}\phi}{dr^{2}_{*}}+\left[\omega^{2}-f\left( m^{2}_{s}+\frac{1}{r^{2}}\left(\lambda^{2}+\frac{k^{2}}{\alpha^{2}_{g}}\right)+\frac{f'}{r}\right)\right]\phi=0,
	\end{equation}
where $\omega,k$ and $\lambda$ are the frequency, the wave number and angular quantum number of the scalar perturbation respectively. The scalar wave equation of the black hole case (\ref{KGeq}) is related to the black string by the following replacement
	\begin{eqnarray}
		-\frac{\Lambda}{3}&\rightarrow&\alpha^{2}_{m},\nonumber \\
		\tau&\rightarrow&\epsilon_{0},\nonumber \\
		l(l+1)&\rightarrow&\lambda^{2}+\frac{k^{2}}{\alpha^{2}_{g}}.\nonumber
	\end{eqnarray}
Therefore, the formula for the quasinormal frequencies in the black string case is simply given by
	\begin{equation}
		\omega_{n}=\kappa_{\chi}\left\lbrace\pm\sqrt{\frac{V_{0}}{\kappa^{2}_{\chi}}-\frac{1}{4}}-\left(n+\frac{1}{2}\right)i\right\rbrace,
	\end{equation}
where
	\begin{equation}
		V_{0}=\frac{\kappa^{2}_{\chi}}{r_{\chi}\gamma+\epsilon_{0}-2Q^{2}/r^{2}_{\chi}}\left[m^{2}_{s}r^{2}_{\chi}+\lambda^{2}+\left(\frac{k}{\alpha_{g}}\right)^{2}\right].
	\end{equation}
The subscript $\chi$ represents an extremal horizon whose surface gravity is positive. Similar to the black hole case, the approximation formula is confirmed by utilizing AIM~(for dS and AdS) and the Spectral Method~(for AdS)~\cite{Burikham:2019fza}. We find that the formula works very well for $\Delta r/r_{\chi}<0.02$, where $\Delta r$ is the difference between the pair of near-extremal horizons.

\section{Conclusions and Discussions}\label{conclude}

In this work, an approximation formula for QNMs of near-extremal black hole in generalized spherically symmetric spacetime is derived by utilizing the P\"{o}schl-Teller potential. Extending the domain of validity of the approximation used in Ref.~\cite{Cardoso:2003sw} from the small universe scenario where the event horizon approaching the cosmic horizon, the approximation formula is shown to be valid for any near-extremal scenarios where any two of the horizons approaching one another.  This includes the scenario where Cauchy horizon approaching the event horizon while remotely separated from the cosmic horizon, i.e., the large universe scenario in the presence of near-extremal black hole.  In the generic near-extremal limits, the Klein-Gordon wave equation takes a simplified form,
	\begin{equation}
		\frac{d^{2}\phi}{dr^{2}_{*}}+\left(\omega^{2}-\frac{V_{0}(r)}{\cosh^{2}(\kappa_{\chi}r_{*})}\right)\phi=0.\nonumber
	\end{equation}
It is well-known that this kind of differential equation has an exact solution with the complex frequencies
	\begin{equation}
		\omega_{n}=\kappa_{\chi}\left\{\pm\sqrt{\frac{V_{0}}{\kappa^{2}_{\chi}}-\frac{1}{4}}-\left(n+\frac{1}{2}\right)i\right\}.\nonumber
	\end{equation}
The above formula can be used to calculate quasinormal modes of near-extremal spherically symmetric black holes and non-rotating black strings for various near-extremal conditions. Numerical calculations by AIM and spectral method are performed to confirm the validity of the approximation formula for asymptotic dS and AdS cases in Ref.~\cite{Burikham:2019fza}.

It is found in Ref.~\cite{Burikham:2017gdm} that there are three kinds of QNMs in the generalized black hole spacetime, near-event-horizon, near-cosmic-horizon, and all-region~(or WKB) modes. The near-horizon modes do not obey Eq.~(\ref{FM}) but are rather given by the formulae $\omega=i\kappa_{\chi}n$~(for non-positive integer $n$) given in Ref.~\cite{Tangphati:2018jdx} when the charge of the scalar is set to zero. From Eqn.~(\ref{KGeq}), for the near-horizon QNMs, the equation of motion can be rewritten as 
	\begin{equation}
		\frac{d^{2}\phi}{dr^{2}_{*}}-\left[\kappa_{\chi}^{2}n^{2}+f(r)\left(m^{2}_{s}+\frac{l(l+1)}{r^{2}}+\frac{f'(r)}{r}\right)\right]\phi=0. \label{nheq}
	\end{equation}
By substitution $\phi(r_{*})\sim e^{ikr_{*}}$, yields $k^{2}\simeq -[...]<0$, where $[...]$ is the bracket term in the above equation. This implies that $k$ must be imaginary in a region where the term in $[..]$ changes slightly in $r_{*}$. Namely, the wave function decays exponentially in the large distance away from the corresponding horizon. In this sense that we say the near-horizon QNMs are concentrated around the horizon and fading away in the far region. On the other hand, the QNMs from the P\"{o}schl-Teller approximation in this work are the all-region modes, their wave functions are not decaying exponentially in the far region and they can reach further distance from the horizons.  

The existence of different kinds of quasinormal frequencies suggests that other than parameters of black hole~(mass, charge and spin), the observed quasinormal frequencies also depend on location of the observer. It would be interesting to explore under what circumstances each type of QNMs will be produced in what proportion and whether we can observe every type of QNMs from actual events of black hole ringdown.

Using the almost-exact formula for QNMs of near-extremal black hole~(becomes more and more accurate as the black hole approaches extremality), we investigate two versions of SCC, the Christodoulou's~(CSCC) and the $C^{1}$-SCC. CSCC states that spacetime cannot be extended through the Cauchy horizon with locally square integrable Christoffel symbols, i.e., finite energy while $C^{1}$-SCC requires that spacetime must be $C^{1}$ inextendible at the Cauchy horizon. The near-horizon QNMs have larger imaginary part giving larger value of $\beta$ (since for nonzero mode with lowest imaginary part, $\omega=-i\kappa_{H}$, they always give $\beta = R = 1$ for extremals and $\beta < 1$ for the near-extremals with $R<1$) than the modes we are considering, so they become redundant since these near-horizon modes just tells us that $C^{1}$-SCC is valid but CSCC might not. It thus requires further investigation into the other modes, namely the all-region modes. The bounds (\ref{CS}) derived from the QNMs formula (\ref{w1}) are proved~(see Appendix \ref{SCCP}) to contain point-wise gaps at $H=1,4,16,36$ in extremal limits which cannot be covered by all possible bounds. 

In the GR-type RN-dS extremal black holes with $\gamma =0$, the CSCC is found to be violated for certain range of BH parameters corresponding to $H=4,16,36$ as listed in Table~\ref{sumtab} and \ref{sumtab1}. At $H=4,16,36$, there are RN-dS extremals that violate CSCC but none of RN-AdS extremal. The BH parameters for each $H=4, 16, 36$ entry in Table~\ref{sumtab1} expand to three contours in the multi-dimensional parameter space when the rescaling in Appendix~\ref{ReS} is applied, i.e., varying $M$. These correspond to three classes of RN-dS extremal BHs that violate CSCC. For near-extremals with $R<1$, the gaps at $H=4,16,36$ are completely closed and only $H=1$ gap remains. This gap, however, corresponds to RN-AdS BH and thus the $\beta$-criteria is not applicable since there is no cosmic horizon. Figure~\ref{SCCzerogam} shows that any near-extremal BHs in GR that have $H=1$ must be RN-AdS with no cosmic horizon. The mass inflation discussed in Ref.~\cite{Luk:2015qja} could potentially destabilize the Cauchy horizon rendering CSCC valid if the resulted singularity is strong. However, if the resulting singularity at the Cauchy horizon is weak, CSCC could be violated~\cite{Bhattacharjee:2016zof}. 

For more generic BH with nonzero $\gamma$ which has no GR counterparts, CSCC can be violated in a number of ranges of BH parameters corresponding to $H=1,4,16,36$ as discussed in Section~\ref{SCCSec} and listed in Table \ref{sumtab1} and \ref{sumtab2}. Each entry of Table~\ref{sumtab1} and \ref{sumtab2} expands to contour in multi-dimensional parameter space when the rescaling in Appendix~\ref{ReS} is applied. They correspond to various classes of BHs with potential CSCC violation categorized by the value of $H$. Notably, we found near-extremal BH with {\it sub-extremal} charge $Q^{2}<M$ that violates CSCC at $H=1$ for near-extremal parameter $r_{H}-r_{C}=10^{-2}$, the near-extremal BH in asymptotically dS with positive $\gamma$ as shown in Table~\ref{sumtab2}. In Table~\ref{sumtab1} for $H=1$, an extremal dS BH with positive $\gamma$ is also found with {\it sub-extremal} charge. On the other hand, near-extremal BHs in asymptotically AdS with $H=1$~(listed in Table~\ref{sumtab2}) have no cosmic horizon and thus the fate of CSCC depends on the tail behaviour of the QNMs in the background which is still an open question. However, CSCC violation is prefered for certain reasonable assumptions of the tail behaviour~\cite{Bhattacharjee:2016zof}.

\acknowledgments

P.B. is supported in part by the Thailand Research Fund (TRF),
Office of Higher Education Commission (OHEC) and Chulalongkorn University under grant RSA6180002.  T.W. is financially supported by Chulalongkorn University, Dutsudi Phiphat Scholarship.

\appendix

\section{Rescaling Parameter for Alternative Value of $M$}\label{ReS}

Throughout this paper the value of $M$ is set to unity. However, it is possible to rescale $M$ to any other values. This can be achieved by rescaling $r=M\tilde{r}$ and rewriting the metric function, spacetime parameters and the Klein-Gordon equation in terms of $\tilde{r}$ and rescaled quantities. These transformations are listed in Table \ref{RP}.
	
\begin{table}[h]
		\centering\
		\begin{tabular}{|c|c|c|}
			\hline
			Quantity&~General Spacetime Parameter~~&~The Tilde Notation~~\\
			\hline
			Mass&$M$&$1$\\
			\hline
			Charge&$Q$&$\tilde{Q}=Q/M$\\
			\hline
			~Graviton Self-interaction parameter~~&$\gamma$&$\tilde{\gamma}=\gamma M$\\
			\hline
			Cosmology Constant&$\Lambda$&$\tilde{\Lambda}=\Lambda M^{2}$\\
			\hline
			Scalar Mass&$m_{s}$&$\tilde{m}_{s}=m_{s}M$\\
			\hline
			Quasinormal Frequency&$\omega$&$\tilde{\omega}=\omega M$\\
			\hline
		\end{tabular}
		\caption{Table of rescaled parameters}
		\label{RP}
	\end{table}

\section{Proof of $R<1$ for $r_{H}\to r_{C}$ near-extremals with existence of cosmic horizon}  \label{Rproof}

In the inspection of SCC violation for near extremal BHs, we need to consider only $r_{H}\gtrsim r_{C}$ case with $r_{\Lambda}>r_{H}$ for dS and $r_{\Lambda+},r_{\Lambda-}>r_{H}$ for AdS respectively.  In both cases, $R$ can be expressed as
\bea
R&&~=\frac{\kappa_{\chi}}{|\kappa_{Cauchy}|}=-\frac{\kappa_{H}}{\kappa_{C}}=\frac{r_{C}^{2}(r_{H}-C)(r_{H}-D)}{r_{H}^{2}(r_{C}-C)(r_{C}-D)}, \notag \\
  &&~=\left( \frac{r_{C}}{r_{H}}\right)\frac{1-\frac{C}{r_{H}}}{1-\frac{C}{r_{C}}}\frac{D-r_{H}}{D-r_{C}}=\left( \frac{r_{C}}{r_{H}}\right)^{2}\frac{C-r_{H}}{C-r_{C}}\frac{D-r_{H}}{D-r_{C}}<1,
\eea
since $\displaystyle{\frac{1-\frac{C}{r_{H}}}{1-\frac{C}{r_{C}}}},\displaystyle{\frac{D-r_{H}}{D-r_{C}}}<1$ for $C<0,D>r_{H}~(C=\text{negative root}, D=r_{\Lambda})$ in the dS case and $\displaystyle{\frac{C-r_{H}}{C-r_{C}}},\displaystyle{\frac{D-r_{H}}{D-r_{C}}}<1$ for $C,D>r_{H}~(C=r_{\Lambda-}, D=r_{\Lambda+})$ in the AdS case respectively.

\section{Proof of SCC bounds}\label{SCCP}

In this section, we will prove that the value $H=1,4,16$ cannot be covered by the lower and upper bounds $\displaystyle{\frac{n^{2}-1/4}{l(l+1)}}\equiv {\rm LB}$ and $\displaystyle{\frac{n(n+1)}{l(l+1)}}\equiv {\rm UB}$ for $R=1$ extremal BHs. We will consider only $n,l>0$ case since for $l=0$, $\beta =n>1/2$ for $n>0$ and thus irrelevant. The constant $c$ below can be negative as long as $n$ is kept positive. 

Starting with assuming $n=l+c, c\in \mathbb{Z}$, the bounds can be rewritten as
\bea
{\rm LB}&&~=\frac{(l+c-\frac{1}{2})(l+c+\frac{1}{2})}{l(l+1)}=\frac{l^{2}+2cl+c^{2}-\frac{1}{4}}{l^{2}+l}, \label{n1l}\\
{\rm UB}&&~=\frac{(l+c)(l+c+1)}{l(l+1)}=\frac{l^{2}+(2c+1)l+c^{2}+c}{l^{2}+l}, \label{n1u}
\eea
For $c\geq 1$, LB and UB are always larger than 1. For $c=0$, LB is always smaller than 1 and UB$=1$. And for $c\leq -1$ since $\displaystyle{\frac{l+c-1/2}{l}}, \displaystyle{\frac{l+c+1/2}{l}}<1$ and $\displaystyle{\frac{l+c}{l}}, \displaystyle{\frac{l+c+1}{l}}<1$, LB and UB are always smaller than 1.  Consequently, the bounds LB~$<H<$~UB never cover $H=1$ but can be made arbitrarily close to $1$ from both sides for sufficiently large $(n,l)$.

Next we rewrite the form of $n$ as $n=2l+c, c\in \mathbb{Z}$, the bounds then become
\bea
{\rm LB}&&~=\frac{(2l+c-\frac{1}{2})(2l+c+\frac{1}{2})}{l(l+1)}=4~ \frac{(l+\frac{c}{2}-\frac{1}{4})(l+\frac{c}{2}+\frac{1}{4})}{l(l+1)}   \notag \\
&&~=4~ \frac{l^{2}+cl+(c/2)^{2}-(1/4)^{2}}{l^{2}+l}, \label{n2l}\\
{\rm UB}&&~=4~ \frac{(l+\frac{c}{2})(l+\frac{c+1}{2})}{l(l+1)}. \label{n2u}
\eea
For $c\geq 1$, LB and UB are always larger than 4. For $c=0$, LB and UB are always smaller than 4. And for $c\leq -1$ since $\displaystyle{\frac{l+c/2-1/4}{l}}, \displaystyle{\frac{l+c/2+1/4}{l}}<1$ and $\displaystyle{\frac{l+c/2}{l}}, \displaystyle{\frac{l+(c+1)/2}{l}}<1$, LB and UB are always smaller than 4.  Consequently, the bounds LB~$<H<$~UB never cover $H=4$ but can be made arbitrarily close to $4$ from both sides for sufficiently large $(n,l)$.

Lastly, by rewriting $n=4l+c, c\in \mathbb{Z}$, the bounds then become
\bea
{\rm LB}&&~=\frac{(4l+c-\frac{1}{2})(4l+c+\frac{1}{2})}{l(l+1)}=16~ \frac{(l+\frac{c}{4}-\frac{1}{8})(l+\frac{c}{4}+\frac{1}{8})}{l(l+1)}   \notag \\
&&~=16~ \frac{l^{2}+cl/2+(c/4)^{2}-(1/8)^{2}}{l^{2}+l}, \label{n4l}\\
{\rm UB}&&~=16~ \frac{(l+\frac{c}{4})(l+\frac{c+1}{4})}{l(l+1)}. \label{n4u}
\eea
For $c = 1$, LB and UB are always smaller than 16 for $l>1/2$. For $c \geq 2$, LB and UB are always {\it larger} than 16. For $c=0$, LB and UB are always smaller than 16. And for $c\leq -1$ since $\displaystyle{\frac{l+c/4-1/8}{l}}, \displaystyle{\frac{l+c/4+1/8}{l}}<1$ and $\displaystyle{\frac{l+c/4}{l}}, \displaystyle{\frac{l+(c+1)/4}{l}}<1$, LB and UB are always smaller than 16.  Consequently, the bounds LB~$<H<$~UB never cover $H=16$ but can be made arbitrarily close to $16$ from both sides for sufficiently large $(n,l)$.

This completes the proof that $H=1,4,16$ can never be covered by the CSCC bounds and thus CSCC is violated at these points. The proof for $H=36$ follows the same argument with $n=6l+c$. Only at $c=2, l=1$ that UB$~=1$, other possibilities all give both LB and UB either smaller or larger than 36 at the same time. As a consequence, $H=36$ is also not covered by the bounds. 

\begin{figure}[h]
		\centering
		\subfloat[]{\includegraphics[width=0.4\textwidth]{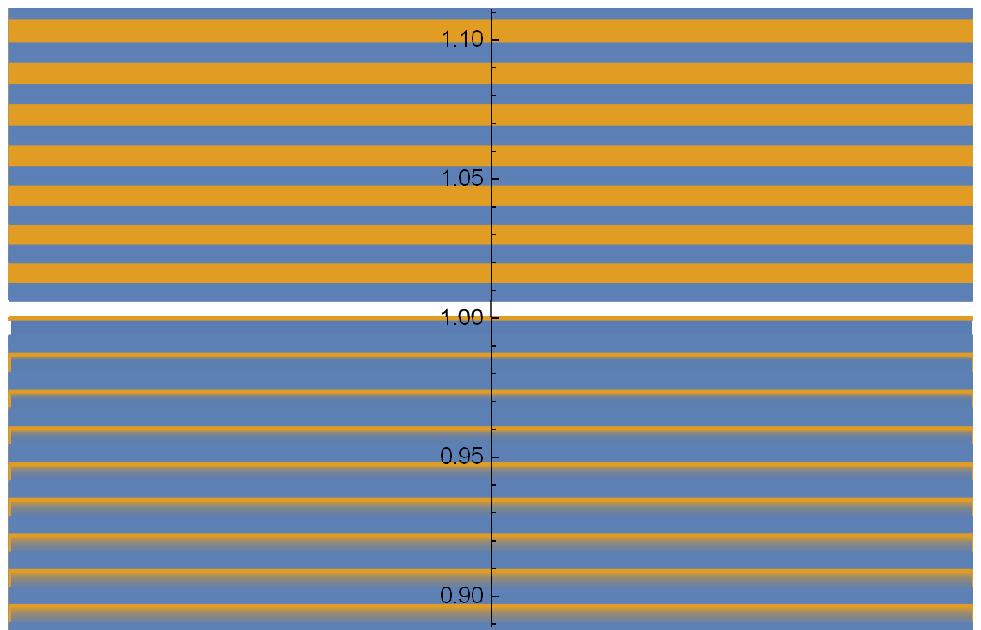}}
		\hspace{1em}
		\subfloat[]{\includegraphics[width=0.4\textwidth]{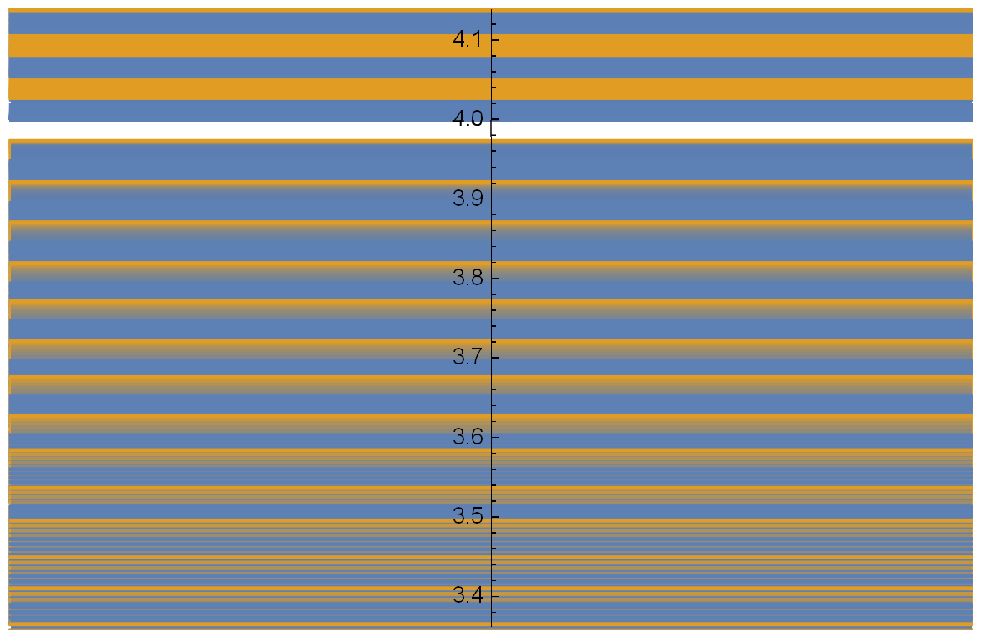}}\\
		\hspace{1em}
		\subfloat[]{\includegraphics[width=0.4\textwidth]{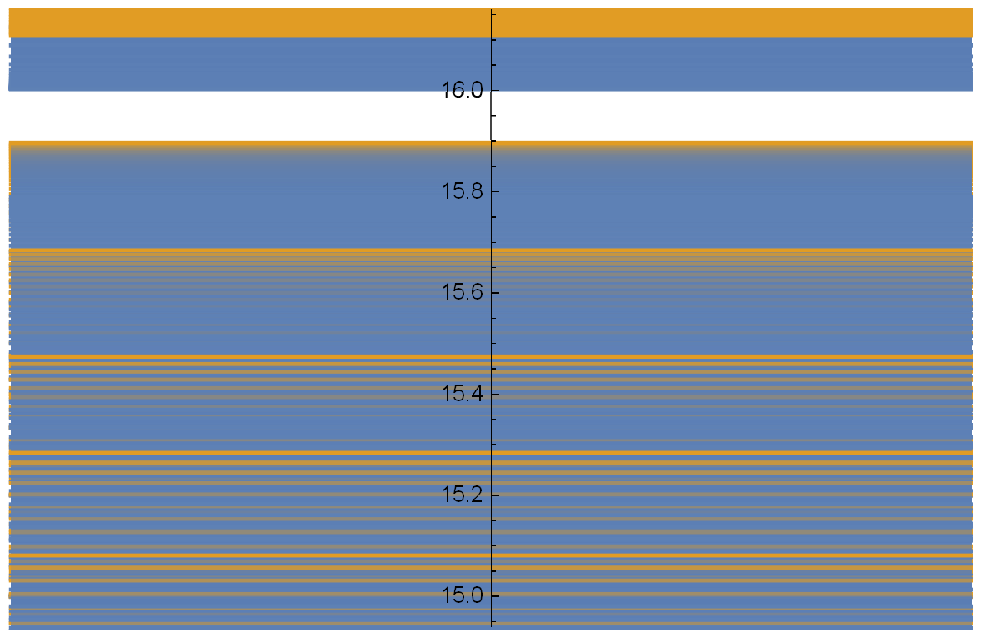}}
		\hspace{1em}
		\subfloat[]{\includegraphics[width=0.4\textwidth]{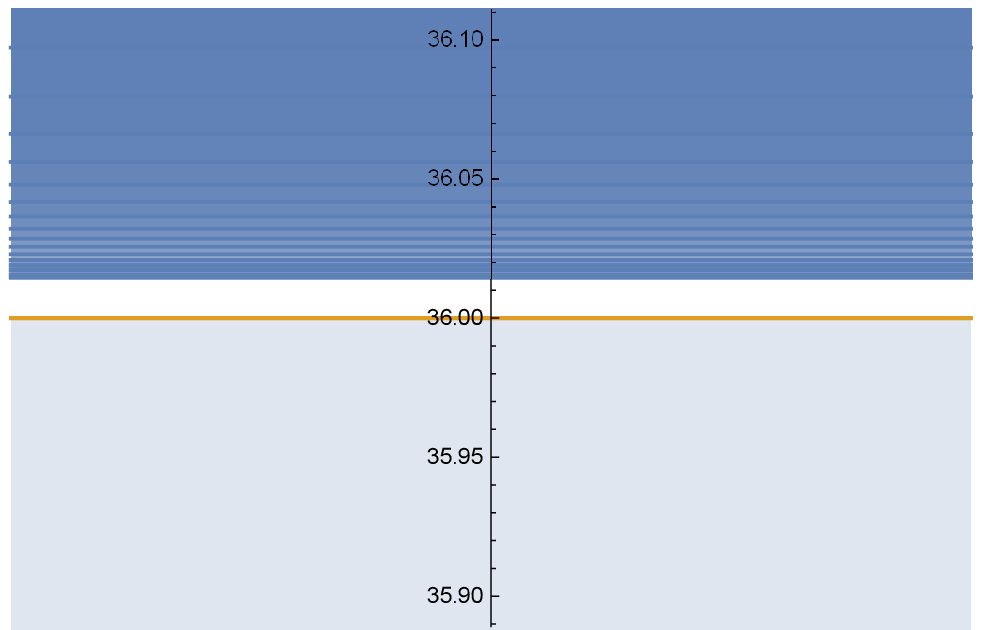}}
		\caption{Covering region of bounds for $n=1-150,l=1-150$, four gaps at 1,4,16,36 are still visible. There are upper bounds locating right at $H=1,36$ but the points $H=1,36$ are not covered by the open bounds. }
		\label{boundsfig}
	\end{figure}

For demonstration, the numerical bounds, LB and UB for $n=1-150,l=1-150$, are shown in Fig.~\ref{boundsfig}. The gaps will be closed up as $n,l$ grow larger.

\end{document}